\documentclass[fleqn,usenatbib]{mnras}

\usepackage{newtxtext,newtxmath}

\usepackage[T1]{fontenc}
\usepackage{ae,aecompl}
\usepackage{lscape}
\usepackage{longtable}
\usepackage{bm}

\usepackage{graphicx}	
\usepackage{amsmath}	
\usepackage{amssymb}	

\title[Morphology-accretion link in FRII-LERGs]{Radio morphology-accretion mode link in FRII low-excitation radio galaxies}

\author[D. Macconi et al.]{D. Macconi,$^{1,2}$\thanks{E-mail: duccio.macconi2@unibo.it}, E. Torresi$^{2}$, P. Grandi$^{2}$, B. Boccardi$^{3}$ and C. Vignali$^{1,4}$\\
$^{1}$Dipartimento di Fisica e Astronomia, Alma Mater Studiorum, Universit\`a degli Studi di Bologna, Via Gobetti 93/2, I-40129 Bologna, Italy \\
$^{2}$INAF-Osservatorio di Astrofisica e Scienza dello Spazio di Bologna, Area della Ricerca CNR, Via Gobetti 101, I-40129 Bologna, Italy\\
$^{3}$Max-Planck-Institut f{\"u}r Radioastronomie, Auf dem H{\"u}gel 69, D-53121 Bonn, Germany\\
$^{4}$INAF-Osservatorio di Astrofisica e Scienza dello Spazio di Bologna, Via Gobetti 93/3, I-40129 Bologna, Italy}

\date{Accepted XXX. Received YYY; in original form ZZZ}

\pubyear{2020}

\begin{document}
\label{firstpage}
\pagerange{\pageref{firstpage}--\pageref{lastpage}}
\maketitle

\begin{abstract}

 Fanaroff-Riley II low-excitation radio galaxies (FRII-LERGs) are characterized by weak nuclear excitation on pc-scales and by properties typical of powerful FRIIs (defined as high-excitation, hereafter HERGs/BLRGs) on kp-scales. Since a link between the accretion properties and the power of the produced jets is expected both from theory and observations, their nature is still debated.
In this work we investigate the X-ray properties of a complete sample of 19 FRII-LERGs belonging to the 3CR catalog, exploiting \textit{Chandra} and XMM-\textit{Newton} archival data. We also analyze 32 FRII-HERGs/BLRGs with \textit{Chandra} data as a control sample. We compared FRII-LERG and FRII-HERG/BLRG X-ray properties 
and optical data available in literature 
to obtain a wide outlook of their behavior.  The low accretion rate estimates for FRII-LERGs, from both X-ray and optical bands, allow us to firmly reject the hypothesis for that they are the highly obscured counterpart of powerful FRII-HERGs/BLRGs.
Therefore, at least two hypothesis can be invoked to explain the FRII-LERGs nature: (i) they are evolving from classical FRIIs because of the depletion of accreting cold gas in the nuclear region, while the extended radio emission is the heritage of a past efficiently accreting activity; (ii) they are an intrinsically distinct class of objects with respect to classical FRIs/FRIIs. Surprisingly, in this direction a correlation between accretion rates and environmental richness is found in our sample. The richer the environment, the more inefficient is the accretion. In this framework, the FRII-LERGs are intermediate between FRIs and FRII-HERGs/BLRGs both in terms of accretion rate and environment. 
  
\end{abstract}



\section{Introduction} 
Black holes (BHs) inhabit the center of almost all massive galaxies and span a wide range of masses, from $10^6~M_\odot$ to several $10^9~M_\odot$, depending strongly on the properties of the host galaxy \citep{Kormendy1993,Magorrian1998,Gebhardt2000,Ferrarese2000,Tremaine2002,Kormendy2004,Greene2005,Greene2006,McConnell2011}.\\ A small fraction of galaxies ($\sim$1\%) show nuclear activity: they are called active galactic nuclei (AGN). AGN produce an enormous amount of energy in a tiny volume ($<<$pc) via gravitational accretion of matter onto the central BH. 
About 10\% of AGN produce strong relativistic jets which emit non-thermal radiation over the (almost) whole electromagnetic spectrum, thus they are defined radio-loud AGN or {\it jetted} AGN (see \citealt{Padovani2016}).
Radio-loud AGN with strongly boosted emission associated to jets and pointing towards the observer's line-of-sight are called blazars. Conversely, radio-loud AGN whose jets are oriented close to the plane of the sky are called Radio Galaxies (RGs).\\ 
RGs were classified for the first time by \cite{Fanaroff1974} following radio morphological criteria. Indeed, Fanaroff-Riley type one objects (FRI) show the highest surface brightness near the core along the jets (i.e. edge-darkened sources), while type two (FRII) objects are characterized by higher surface brightness at the lobes extremities, far from the nucleus (i.e. edge-brightened). There are some cases for which the morphological classification is ambiguous, e.g. they show FRII-like jets on one side and FRI-like on the other: the so-called hybrid double sources \citep{Gopal-Krishna2000}. \\ 
The FRI-FRII morphological dichotomy quite neatly translates into a separation in terms of extended radio power: sources with radio luminosity at 178~MHz below $10^{26}$ W~Hz$^{-1}$ \citep{Fanaroff1974,Tadhunter2016} are generally FRI, while sources with luminosity above this threshold tipically belong to FRII class.\\

RGs can be also classified on the basis of their optical spectra, accordingly to the emission lines produced in the Narrow Line Regions (hereafter NLR; \citealt{Laing1994,Buttiglione2009,Buttiglione2010,Buttiglione2011}), in High-Excitation or Low-Excitation Radio Galaxies (HERGs and LERGs, respectively). The main diagnostic of high excitation is the [OIII]$\lambda$5007 line luminosity. Conversely, the standard spectroscopic indicators of low excitation are [NII]$\lambda$6584, [SII]$\lambda$6716 and [OI]$\lambda$6364 (\citealt{Buttiglione2010}). To take into account simultaneously both high- and low-excitation diagnostics,  \cite{Buttiglione2010} introduced the exctitation index (EI).\footnote{The excitation index can be defined as:\\ {\tiny $EI=log~[OIII]/H_{\beta}~-~1/3~(log~[NII]/H_{\alpha}~+~log~[SII]/H_{\alpha}~+~log~[OI]/H_{\alpha})$ .}}
Considering that  different excitation modes of the NLR are associated to different accretion rates \citep{Gendre2013,Heckman2014}, this spectroscopic classification reflects the accretion regime at work in the central regions of the AGN.  In particular, HERGs accrete efficiently (quasar-mode, L/L$_{\rm Edd}>$0.1), i.e. the potential energy of the gas accreted by the super-massive black hole (SMBH) is efficiently converted into radiation \citep{adaf}. Conversely, LERGs are characterized by low accretion rates typical of radiatively inefficient hot accretion flows (jet-mode, L/L$_{\rm Edd}$ $\leq$0.01-0.1), and the jet carries the bulk of the AGN energy output \citep{adaf, Heckman2014}.\\
Over the years, studies on radio-loud AGN were fundamental to investigate the jet-launching mechanism and the eventual connection between ejection of relativistic jets and accretion of material onto the central BH. 
Indeed, jets are produced close to the BH and their power is predicted to depend on the BH properties (mass and spin) and on the magnetic field strength: in \cite{Blandford1977} model, the jet power ($P_{jet}$) is proportional to $(aMB)^2$, where $a$ is the BH spin, $M$ its mass and $B$ is the magnetic field value at the BH horizon \citep{Ghisellini2014}.\\
In any case, the magnetic field plays in general a major role in channeling power from the BH \citep{Blandford1977}, or from the disk \citep{Blandford1982}, into the jet \citep{Maraschi2003}.
In both scenarios, if the magnetic field strength depends, as generally assumed, on the accretion rate (see e.g. \citealt{Ghisellini2014}), a relation between the accretion rate and the jet power is expected. Various observational studies seem to confirm the link between accretion and ejection (e.g.: \citealt{Rawlings1991,Celotti1997,Willott1999,Maraschi2003,Ghisellini2010,Ghisellini2014}).\\ 
In the general picture, FRII radio galaxies host an efficient accretion disk \citep{Shakura1973}, while FRIs are characterized by hot inefficient accretion flows (ADAF-like; \citealt{Narayan1994,Abramowicz1995}). 
However, there is a group of FRII sources that does not fit into this framework. They exhibit powerful extended radio structures but inefficient accretion, attested by their optical spectra typical of LERGs. This kind of objects is not so exotic: indeed, about 25\% of sources belonging to the 3CR catalog at $z<0.3$ and having both radio and optical classifications, are FRII-LERGs \citep{Buttiglione2009,Buttiglione2010,Buttiglione2011}. \\
Therefore, given their peculiar nature  and not negligible number, FRII-LERGs constitute a particularly relevant population for the comprehension of the role of the central engine in powering RG jets and in shaping the extended radio morphology.\\
Within this context, the X-ray band is a fundamental tool to probe the processes at work on different scales from sub-pc up to several hundreds of kpc.
To this aim, we performed a systematic  and homogeneous X-ray analysis of all FRIIs belonging to the 3CR sample \citep{Bennett1962}  below z<0.3, one of the best studied radio catalogs in all energy bands.
 For the first time FRII-LERGs are explored as a separate class. Their X-ray and multi-frequency properties are compared to those of FRII-HERGs, in order to understand their nature. For example, they could be FRII-HERGs seen through a thicker obscuring screen or they could have central engines in a ``switching-off" phase, in which the standard accretion flow becomes inefficient (ADAF-like).
Finally, also the role played by the environment in triggering the AGN and the link between environment and jet power are explored. \\
This work is organized as follows: in \S 2 we describe the sample of sources. In \S 3 we report on the X-ray data reduction and analysis, and discuss our results  in  \S 4. In \S 5 we summarize our conclusions.   
Throughout the paper we adopt: $H_0=71$~km~s$^{-1}$~Mpc$^{-1}$, $\Lambda_\Omega=0.73$, $\Lambda_m=0.27$ \citep{Komatsu2009}.

\section{The analyzed sample}
The 3CR sources at $z<0.3$ classified both in the optical (HERGs vs LERGs) and radio bands (FRIs vs FRIIs) are 79. Following \cite{Buttiglione2009,Buttiglione2010,Buttiglione2011}, who provided  the classification in both bands, radio galaxies are divided into:
\begin{itemize}
    \item 30 FRII-HERGs and 17 FRII-BLRGs\footnote{Broad-Lines Radio Galaxies (BLRGs) are classified as HERGs according to their NLR emission. They differ from HERGs for the presence of broad permitted lines in the optical spectrum, coming from the broad line regions (BLR). Therefore, HERGs are classified as type 2 AGN, i.e.  edge on, while BLRGs are type 1s,  i.e. face on.};
   \item 19 FRII-LERGs;
    \item 13 FRIs.
\end{itemize}
 \noindent 

\begin{figure}
   \centering
   \includegraphics[width=7cm] {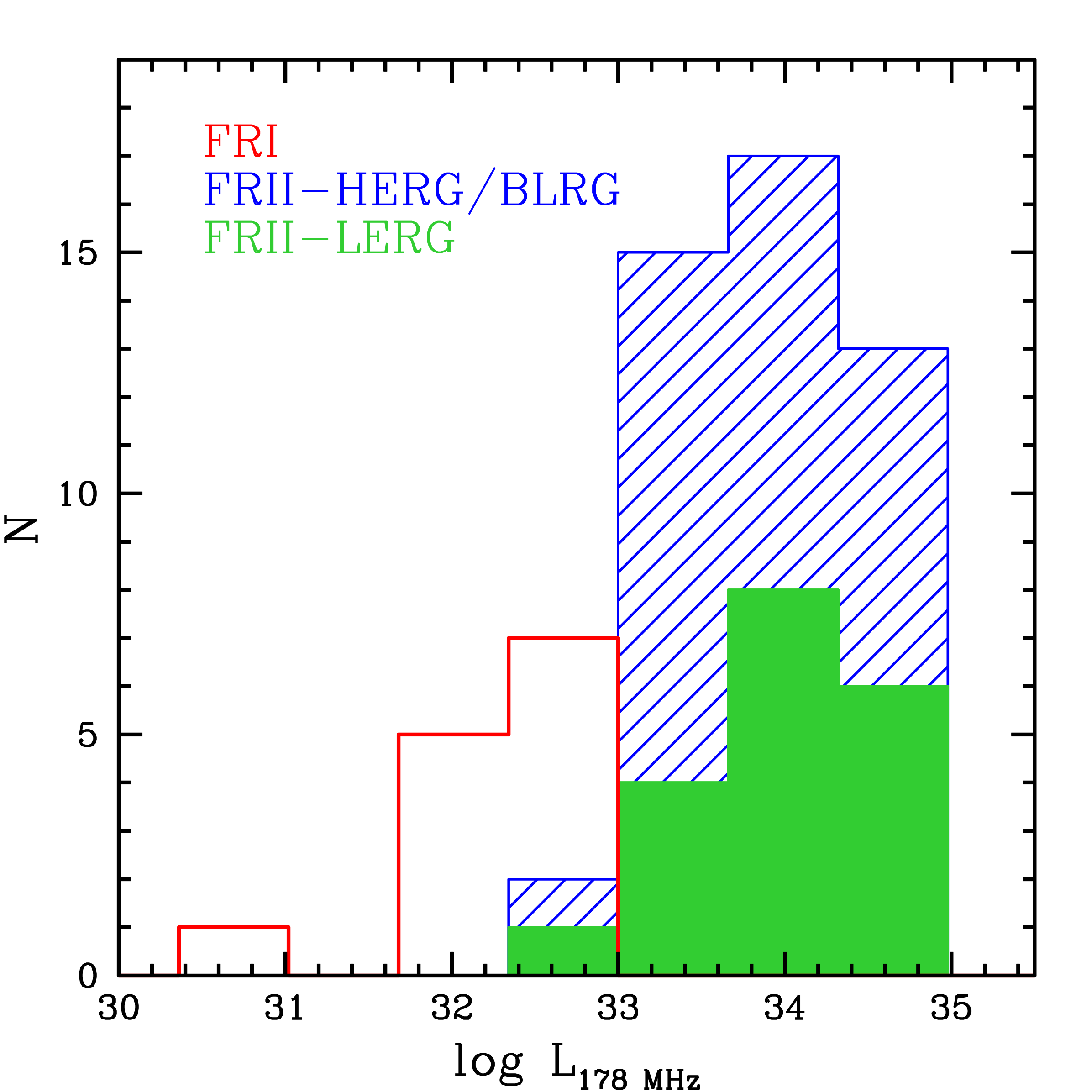}
    \caption{Distribution of total radio luminosity at 178~MHz in units of  erg~s$^{-1}$~Hz$^{-1}$ from \protect\cite{Spinrad1985}: FRII-HERGs/BLRGs are in blue, FRII-LERGs are in green and FRIs are in red. While a separation between FRIs and FRIIs is clear, FRII-HERGs/BLRGs and FRII-LERGs completely overlap.}\label{Pradio}
 \end{figure}
From the radio point of view, the majority of the sources are powerful FRIIs (66), distributed over the entire redshift range, while FRIs (13) represent  a small fraction of the total ($\sim16\%$) and cluster below z<0.05.
This is not surprising since low-power RGs are known to be mainly observed  at lower redshifts \citep{Laing1983,Spinrad1985,Wall1985,Morganti1993,Burgess2006A,Burgess2006B}. 
From the optical point of view, the number of  HERGs/BLRGs (47) and LERGs (32) is not very different, implying that about half of  the local radio sources is characterized by a low accretion regime, independently on the radio classification. Indeed, FRII-LERGs and FRII-HERGs/BLRGs span similar values of radio luminosity at 178 MHz, as shown in Figure \ref{Pradio}. Following the prescriptions of \cite{Willott1999}, if we assume $L_{\rm 178~MHz}$ as a crude signature of the average jet power over time, FRII-LERGs do not fit in the stardand picture linking powerful jets to efficient accretion flows. Other ingredients have to be considered to explain the observations.

\section{X-ray Data Analysis}
\label{data-reduction}

We performed a systematic X-ray analysis of all the FRIIs of the 3CR sample at z<0.3 exploiting archival data from the \textit{Chandra} satellite. This is the most suited telescope to perform such kind of analysis given the good angular resolution ($\approx 0.5''$ in the observational energy band) and the presence of all the sources of the analyzed sample in the public archive \footnote{https://cda.harvard.edu/chaser/ .}.
 Several of them belong to the 3C \textit{Chandra} legacy survey \citep{Massaro2010,Massaro2012}. However, \citeauthor{Massaro2010} adopted a flux sky map method to present the data, while 
here we follow a different approach based on the direct fit of the data.

We consider FRII-LERGs as key targets of our work (main sample). We also analyze X-ray data of FRII-HERGs/BLRGs as a ``control" sample. 
The goal of our analysis is to investigate the nuclear activity of sources down to the innermost regions (sub-pc scales) both in terms of gas presence ($N_{\rm HX}$) and X-ray luminosity.


In Table \ref{tab-log} the observation log is reported. 
All the sources were pointed by \textit{Chandra}.

When more than one observation was available,  the data were combined in order to achieve better statistics.
We also analyzed all the FRII-LERGs belonging to the XMM-\textit{Newton} archive. In three cases, i.e. 3C~349, 3C~353 and 3C~460, we used XMM-\textit{Newton} data since its larger effective area guaranteed a better constraint of the spectral parameters.

\textit{Chandra} data were reprocessed using the software {\sc CIAO} (Chandra Interactive Analysis of Observations), version 4.10 with calibration database CALDB version 4.8.1 and following standard procedures.\\
A preliminary check of the images was necessary to investigate the presence of extended emission. In case the source was extended, two images were produced: a soft one (0.3-4~keV) and a hard one (4-7~keV). This approach helps in better defining the pointlike emission from the core and modeling its spectrum.
Subsequently, VLA radio contours at 1.4~GHz or 5~GHz (with a few arcsec as spatial resolution) were superimposed on the hard X-ray image to properly identify the peak of the nuclear emission and verify the presence of other features such as lobes, jets, knots along the jet.  Given the coincidence between the radio core and the X-ray peak in all the analyzed images, no astrometrical correction was applied.
The core spectrum was usually extracted from circular regions with a radius ranging between 1.5''-2.5'' (depending on the presence or not of extended emission) in order to collect more than 90$\%$ of photons. The background was extracted in a clean circular region in the same CCD of the source, avoiding any contamination from field sources or from the source itself. \\
Spectra were grouped to a minimum of 20 counts per bin in order to adopt the $\chi^2$ statistics. When this was not possible, the C-statistics was applied \citep{Cash1979} and spectra were grouped to at least one count per bin.
For the sources 3C~136.1, 3C~153 and 3C~430 the small number of counts (about 10 counts over the entire spectrum) prevented any modeling. Fluxes and luminosities were estimated using the \textit{Chandra} Proposal Planning Toolkit  (PIMMS)\footnote{\url{http://cxc.harvard.edu/toolkit/pimms.jsp} .} and assuming a simple power-law model with $\Gamma=1.7$ \citep{Grandi2006}.
In 3C~196.1, 3C~288, 3C~310 and 3C~388 the AGN emission is completely overwhelmed by the cluster, thus precluding any nuclear study.

\begin{table}
\footnotesize
\setlength{\arrayrulewidth}{0.3mm}
\setlength{\tabcolsep}{6pt}
\renewcommand{\arraystretch}{0.8}
\begin{center}
\begin{tabular}{p{0.8cm}p{0.06cm}p{1.2cm}p{1.4cm}p{1cm}p{0.5cm}}
\hline
\multicolumn{1}{c}{\textbf{3CR}} & \multicolumn{1}{c}{\textbf{Teles.}} & \multicolumn{1}{c}{\textbf{obsID}} & \multicolumn{1}{c}{\textbf{date}} & \multicolumn{1}{c}{\textbf{CCD}} & \multicolumn{1}{c}{\bm{$t_{exp}$} \textbf{(s)}}\\
\multicolumn{1}{c}{(1)}&\multicolumn{1}{c}{(2)}&\multicolumn{1}{c}{(3)}&\multicolumn{1}{c}{(4)}&\multicolumn{1}{c}{(5)}&\multicolumn{1}{c}{(6)}\\\hline
\multicolumn{6}{c}{\textbf{FRII-LERGs - Main sample}}\\\hline
3C88$^{}$&C&9391&2008-06-30&ACIS-I&11270\\
$^{}$&C&11751&2009-10-14&ACIS-S&20180\\
$^{}$&C&11977&2009-10-06&ACIS-S&50280\\
$^{}$&C&12007&2009-10-15&ACIS-S&35080\\
3C132$^{}$&C&9329&2008-03-26&ACIS-S&7790\\
3C153$^{}$&C&9302&2007-12-07&ACIS-S&8170\\
3C165$^{}$&C&9303&2008-02-02&ACIS-S&7770\\
3C166$^{}$&C&12727&2010-11-29&ACIS-S&8050\\
3C173.1$^{}$&C&3053&2002-11-06&ACIS-S&24310\\
3C196.1$^{}$&C&12729&2011-02-11&ACIS-S&8050\\
3C213.1$^{}$&C&9307&2008-04-14&ACIS-S&8170\\
3C236$^{}$&C&10249&2009-01-14&ACIS-I&41040\\
3C288$^{}$&C&9275&2008-04-13&ACIS-S&40150\\
3C310$^{}$&C&11845&2010-04-09&ACIS-S&58320\\
3C326$^{}$&C&10908&2009-05-10&ACIS-I&27880\\
3C349$^{}$&X&0501620301&2007-08-07& EPIC/pn &14863\\
$^{}$&X&0501621601&2007-10-03& EPIC/pn &15113\\
3C353$^{}$&X&0400930101&2006-08-25& EPIC/pn &44264\\
$^{}$&X&0400930201&2007-02-17& EPIC/pn &10916\\
3C357$^{}$&C&12738&2010-10-31&ACIS-S&8050\\
3C388$^{}$&C&5295&2004-01-29&ACIS-I&31120\\
3C401$^{}$&C&4370&2002-09-21&ACIS-S&25170\\
&C&3083&2002-09-20&ACIS-S&22960\\
3C430$^{}$&C&12744&2011-11-14&ACIS-S&8050\\
3C460$^{}$&X&0675400101&2011-12-24& EPIC/pn&48744\\\hline
\multicolumn{6}{c}{\textbf{FRII-HERGs/BLRGs - Control sample}}\\\hline
3C20&C&9294&2007-12-31&ACIS-S&8040\\
3C33&C&7200&2005-11-12&ACIS-S&20180\\
3C61.1&C&9297&2008-12-05&ACIS-S&8160\\
3C79&C&12723&2010-11-01&ACIS-S&7790\\
3C98&C&10234&2008-12-24&ACIS-I&32130\\
3C105&C&9299&2007-12-17&ACIS-S&8180\\
3C133&C&9300&2008-04-07&ACIS-S&8140\\
3C135&C&9301&2008-01-10&ACIS-S&8040\\
3C136.1&C&9326&2008-01-10&ACIS-S&10040\\
3C171&C&10303&2009-01-08&ACIS-S&60220\\
&C&9304&2007-12-22&ACIS-S&8040\\
3C180&C&12728&2010-12-24&ACIS-S&8060\\
3C184.1 &C&9305   & 2008-03-27 & ACIS-S & 8130\\
3C192  &C&   9270&    	2007-12-18&	ACIS-S&10150\\
 &C&19496&2017-12-18&	ACIS-S&70110\\
 &C&20888&2017-12-21&	ACIS-S&10070\\	 
 &C&20889&2017-12-21& ACIS-S&33110\\	 
 &C&20890&2017-12-24& ACIS-S&21410\\	 
 &C&20891&2017-12-22& ACIS-S&35760\\	 
3C223&C&12731&2012-01-07&ACIS-S&8050\\
3C223.1&C&9308&2008-01-16&ACIS-S&8030\\
3C234&C&12732&2011-01-19&ACIS-S&8050\\
3C277.3&C&11391& 2010-03-03&ACIS-S&25120\\
&C&15023& 2014-03-15&ACIS-I&44080\\
&C&15024& 2014-03-16&ACIS-I&20090\\
&C&16600& 2014-03-11&ACIS-I&98080\\
&C&16599&  2014-03-13&ACIS-I&29090\\
3C284&C&12735&2010-11-17&ACIS-S&8050\\
3C285&C&6911&2006-03-18&ACIS-S&40150\\
3C300&C&9311&2008-03-21&ACIS-S&8040\\
3C303.1&C&9312&2008-02-21&ACIS-S&7770\\
3C305&C&9330&2008-04-07&ACIS-S&8330\\
&C&12797&2011-01-03&ACIS-S&29040\\
&C&13211&2011-01-06&ACIS-S&29040\\
3C321&C&3138&2002-04-30&ACIS-S&47730\\
3C327&C&6841&2006-04-26&ACIS-S&40180\\
3C379.1&C&12739&2011-04-04&ACIS-S&8050\\
3C381&C&9317&2008-02-21&ACIS-S&8170\\
3C403&C&2968&2002-12-07&ACIS-S&50130\\
3C436&C&9318&2008-01-08&ACIS-S&8140\\
&C&12745&2011-05-27&ACIS-S&8060\\
3C452&C&2195&2001-08-21&ACIS-S&80950\\
3C456&C&12746&2011-01-17&ACIS-S&8050\\
3C458&C&12747&2010-10-10&ACIS-S&8050\\
3C459   &C&12734  & 2011-10-13 & ACIS-S & 8050\\
   &C&16044  & 2014-10-12 & ACIS-S & 59960\\\hline
  \end{tabular}
\normalsize
\caption{Observation log of FRII-LERGs (main sample) and FRII-HERGs/BLRGs (control sample). Column description: (1) 3CR name; (2) Telescope: C=\textit{Chandra} and X=XMM-\textit{Newton}; (3) Observation ID; (4) Start date of the observation; (5) Instrument used in the observation; (6) Total exposure time in seconds. All the sources are the target of the observation.}\label{tab-log}
\end{center}
\end{table}

XMM-\textit{Newton} data were reduced using the Scientific Analysis Software (SAS) version 16.1 together with the latest calibration files and following standard procedures. Throughout the paper results refer to EPIC/pn data, but all the EPIC instruments were checked.\\
Source and background spectra were extracted in 0.5-10 keV band from circular regions with radius varying between 20'' and 30'', depending on the source extension, in order to maximize the S/N ratio. In all cases, at least 80$\%$ of photons fell within the extraction region. The background was chosen in a circular region in the same CCD of the source, avoiding any contamination from field sources or from the source itself.\\
Spectra were grouped to a minimum of 20 counts per bin and the $\chi^{2}$ statistics was applied.

We checked for the presence of pile-up effects in each source (using the PIMMS software for \textit{Chandra} data and the task {\sc EPATPLOT} in the SAS for  \textit{XMM-Newton} data). The pile-up was generally negligible ($<10\%$) or absent in FRII-LERGs, but turned out to be important in FRII-HERGs/BLRGs seen face-on (i.e. BLRGs). Indeed, we could perform a \textit{Chandra} spectral analysis only for two BLRGs, i.e. 3C~184.1 and 3C~459,  for which the estimated pile-up  was less than $<10\%$.

The spectral analysis was performed using the XSPEC version 12.9.1 \citep{Arnaud1996}.
The energy range considered in the spectral fitting was 0.3-7 keV for \textit{Chandra} and 0.5-10 keV for XMM-\textit{Newton}. Errors reported are quoted at 90\% confidence for one parameter of interest \citep{Avni1976}.

\subsection{Spectral analysis}
\label{spectral-analysis}
\begin{figure*}
\centering
\includegraphics[width=7cm,height=6cm]{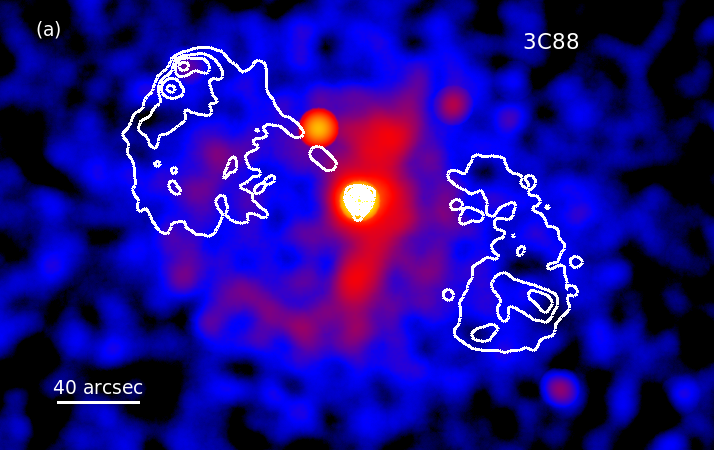}
\hspace{0.1cm}
\includegraphics[width=7cm,height=6cm]{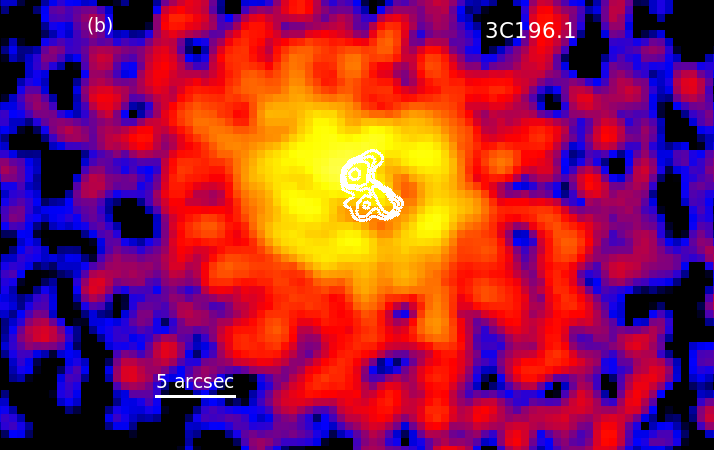}
\hspace{0.1cm}
\includegraphics[width=7cm,height=6cm]{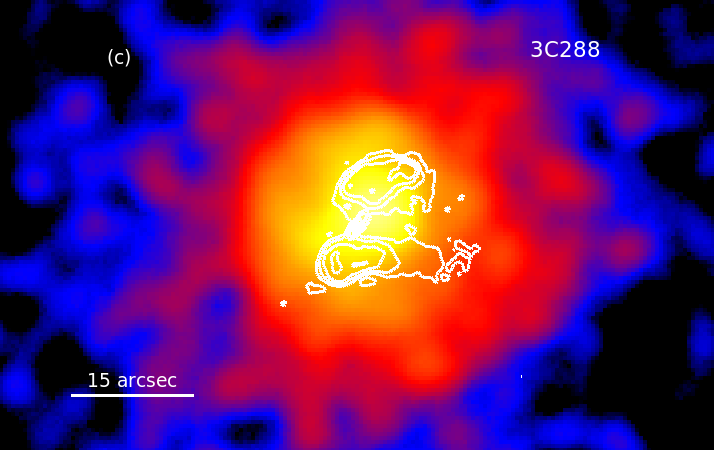}
\hspace{0.1cm}
\includegraphics[width=7cm,height=6cm]{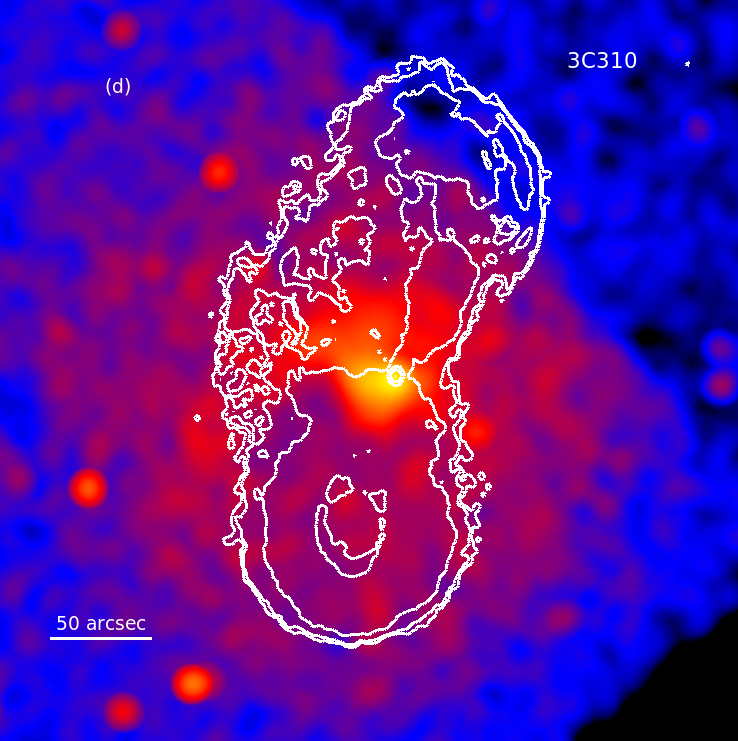}
\hspace{0.1cm}
\includegraphics[width=7cm,height=6cm]{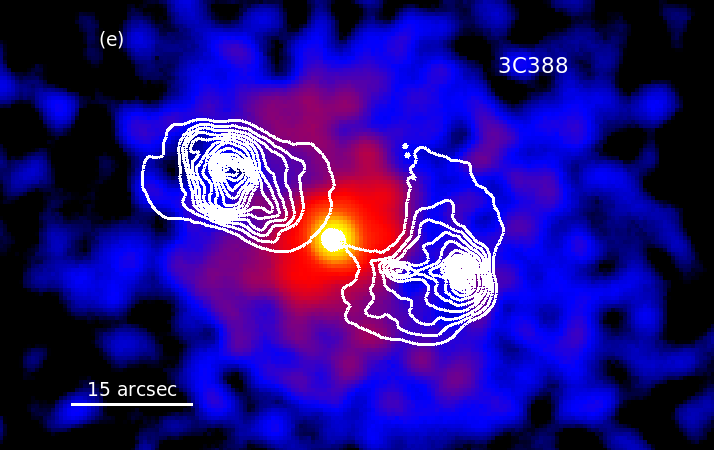}
\hspace{0.1cm}
\includegraphics[width=7cm,height=6cm]{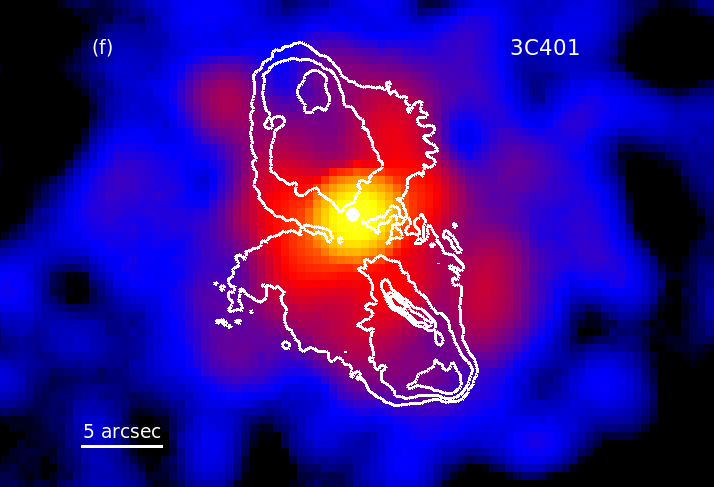}
\caption{\textit{Chandra} 0.3-7~keV images of extended FRII-LERGs. Radio VLA contours at 5~GHz and 1.4~GHz (\textit{white}) are  superimposed to the X-ray images. Panel (a): {\bf 3C~88} inhabits the center of a galaxy group and produce the largest X-ray cavities ever found in such an environment \protect\citep{Liu2019}. Panel (b): {\bf 3C~196.1} is the Brightest Cluster Galaxy (BCG) of a Cool Core Cluster (CCC). \protect\cite{Ricci2018} measured a core cluster temperature kT$\sim$ 3~keV, in agreement with our analysis (see Table \ref{soft_table}). Panel (c): {\bf 3C~288} resides the center of a poor not-CC cluster, as shown by \protect\cite{Lal2010}. They found an ICM gas temperature kT$\sim$3~keV extending up to 400~kpc. Panel (d): {\bf 3C~310} is the central galaxy of a poor cluster  with a temperature of about 3 keV at a distance between 100 and 180 kpc (at which a shock occurs) \protect\citep{Kraft2012}. Panel (e): {\bf 3C~388} resides in the center of a small cluster environment with an ICM temperature of about 3.5~keV and a cool core probably heated by a nuclear outburst \protect\citep{Kraft2006}. 
Panel (f): {\bf 3C~401} is in the center of a cluster with ICM mean temperature of 2.9~keV. \protect\cite{Reynolds2005} proposed both a thermal hot core ($T\approx4.9$~keV) and a simple power law model (which is the model assumed in this work) because statistically indistinguishable.}\label{fig-cluster}
\end{figure*}

An ispection of the X-ray images indicate that 6 out of 19 FRII-LERGs show strong emission over the galaxy-scale (from several tens to hundreds of kpc) due to hot gas from cluster (Figure \ref{fig-cluster}).  
On the contrary, no FRII-HERGs/BLRGs show cluster emission in the \textit{Chandra} images, although some of them show resolved emission on kpc-scale.

At first, we considered as baseline model a single power-law convolved with Galactic column density (\cite{Kalberla2005}; {\sc PHABS}). When the power-law spectral slope was less than 1 an intrinsic absorption component ({\sc ZPHABS}) was added to the fit. Because of the poor statistics and/or the complexity of the emission, we were forced to fix the hard photon index ($\Gamma=1.7$) in 7 out of 19 FRII-LERGs and in 27 out of 32 FRII-HERGs/BLRGs (see Table \ref{tabellone}). If no XMM-\textit{Newton/Chandra} information on the power-law spectral slope was available in literature, we chose a reliable value of $\Gamma=1.7$.
Nonetheless, we checked whether different values of the photon indices could produce significant changes in the estimate of the column density. However, even assuming  different photon indices ($\Gamma=1.4$ and $\Gamma=2.0$), the column densities  and the intrinsic luminosity remain consistent within the errors\footnote{For all the sources with fixed spectral slope, the N$_H$ and L$_X$ values do not change respectively more than $\approx$15\% and 40\% varying the power law spectral slope from 1.4 to 2.0. As the uncertanties of the same quantities in Table \ref{tabellone} are above 70\%, we are confident that the $\Gamma=1.7$ assumption does not significantly affect our results.}.
If residuals were still present at soft-energies a second power-law, or a thermal emission ({\sc MEKAL}) were added to the fit. A second power-law is expected if the primary component is scattered by clouds of electrons above the torus. A thermal emission is expected if the source is embedded in a gaseous environment, i.e. hot corona of early-type galaxies \citep{Fabbiano1992} or intergalactic medium. The MEKAL model could also roughly mimic features related to photoionized gas, given the limited energy resolution of CCD detectors. Therefore, after testing for collisional gas presence, if  prominent photoionized features were present in the soft X-ray spectrum yet, a fit with multiple narrow emission lines (Gaussian-profile) was tested.\\
In the hard spectrum, a Gaussian component ({\sc ZGAUSS}) was included  if  positive residuals were observed in the region of the iron K$\alpha$ line (5-7~keV). 
Once the Fe~k$\alpha$ line was attested, the presence of a reflection component ({\sc PEXRAV}) was verified: in fact, this is expected when cold matter surrounding the nuclear engine reprocesses the primary X-ray radiation \citep{Lightman88}. 
In this case, the cut-off energy and the angle between the normal to the disk and the observer were  fixed to 100~keV and 30$^\circ$, respectively. The reflection component is modeled by the parameter $R=\Omega/2\pi$, corresponding to the
solid angle fraction of a neutral, plane parallel slab
illuminated by the continuum power-law ({\sc PEXRAV}). Given the low statististics and the limited energy range covered by \textit{Chandra} and \textit{XMM-Newton}, small variations in these parameters do not impact the fit.

\subsection{Results}
The results of the X-ray analysis,  listed in Table \ref{tabellone}, are in substantial agreement with those reported in the literature using different satellites and/or different approaches (e.g. \citealt{Grandi2006,Evans2006,Massaro2010,Massaro2012}).
Details on the soft X-ray component and reprocessed features are listed in Tables \ref{soft_table} and \ref{reprocessed_table}, respectively.\\

The photon index of FRII-LERGs was tightly constrained for 8 out of 19 sources: the mean $\Gamma$ value is $1.7$ and the standard deviation is 0.3 (see Table \ref{tabellone}). Intrinsic cold gas obscuration was required in about 50\% of the sources. They are generally moderately absorbed, with a  $N_{\rm HX}$ of the order of a few $10^{22}$~cm$^{-2}$. 
Only in two radio galaxies, i.e. 3C~173.1 and 3C~460, the column density reaches values of few $10^{23}$~cm$^{-2}$.
An iron K$\alpha$ line was detected in 3C~353 (see Table \ref{reprocessed_table})  with an intensity, within the large uncertainties, compatible with being produced by the same matter obscuring the nuclear region \citep{Ghisellini1994}. For the other objects with intrinsic absorption, the feature could not be revealed because of the low statistics and the abrupt drop of the \textit{Chandra} effective area above 6-7~keV. \\
When present, the soft X-ray excess is well described by a power-law, that is probably scattered nuclear emission. Indeed, the normalization values of the scattered component at 1~keV are always a few percent of the absorbed one: the mean value is 6\%, in agreement with those measured for type 2 Seyferts (e.g. \citealt{Bianchi2007}).\\ 
The cluster emission, when present in the X-ray images, is generally dominant. In four cases (3C~196.1, 3C~288, 3C~310, 3C~388), the AGN is overwhelmed by the thermal gas and any nuclear study is precluded. Therefore, for these sources the estimated 2-10~keV luminosity should be considered as an upper limit of the nuclear AGN emission. Only in 3C~88, the AGN spectrum was disentangled from the thermal emission and it is analogous to that of the other absorbed radio galaxies. Instead, in 3C~401 the AGN emission dominates over the cluster one  and the nuclear spectrum is well reproduced by a single power-law. The intrinsic absorption is negligible, and indeed only an upper limit is provided (see Table \ref{tabellone}).\\

\noindent
The spectra of the control sample (FRII-HERGs/BLRGs) are generally more complex than FRII-LERGs (see Table \ref{tabellone}). About 90\% of them show strong obscuration, with typical values one order of magnitude higher than FRII-LERGs ($N_{\rm H}\sim10^{23}$~cm$^{-2}$).  The photon index could be well constrained in only 15\% (5 out of 32) of sources ($<\Gamma>=1.8$ and $\sigma_{rms}=0.5$).
Intense iron lines with Equivalent Width (EW)  spanning from 140~eV to more than 1~keV are detected in 11 sources and, in at least two cases, a Compton reflection model was also required (Table \ref{reprocessed_table}).
These reprocessed features, signature of a complex and inhomogeneous circumnuclear absorber,  are commonly observed in Seyfert-like spectra \citep{Risaliti2002}.

The soft X-ray excess of $>$50\%  of FRII-HERGs/BLRGs is generally well reproduced by a second power-law, which can be interpreted as the scattered component of the primary one.
The mean unabsorbed normalization at 1~keV is 8\% of the absorbed one. 
In addition to the second power-law, a {\sc MEKAL} model is  required in a few objects: in some cases, this component is  directly related to collisional gas emission (cluster or shocked gas), in the other ones it could mimic photoionized  features \citep{Balmaverde2012}. Indeed, in two radio galaxies, i.e. 3C~403 and 3C~321, single soft X-ray emission lines associated to Ne~IX, O~VII and Mg~XI were revealed in the spectrum.\\


\noindent
In summary, the control sample show more complex and feature-rich spectra than the key sample. FRII-HERGs/BLRGs are  characterized by mean values of intrinsic absorption and X-ray luminosity one order of magnitude larger than FRII-LERGs, implying a substantially higher activity of the central engine and more variegated circumnuclear environment.

\begin{table*}
\setlength{\arrayrulewidth}{0.3mm}
\setlength{\tabcolsep}{12pt}
\begin{center}
\caption{Spectral parameters of the X-ray continuum.}
\begin{tabular}{llllllll}
\hline
Name & z & $\rm N_{H,Gal}$     & Fitted  & $\rm N_{H,intr}$ & $\Gamma_{\rm H}$ & $L_{2-10~\rm keV}$ & Statistics$^b$ \\ 
&&($10^{20}~\rm cm^{-2}$) &  Model$^a$     &($10^{22}$~cm$^{-2}$) & &(10$^{42}$~erg~s$^{-1}$) &\\
\hline
\multicolumn{8}{c}{\textbf{FRII-LERGs}}\\

 3C~88              & 0.0302  & 8.3  & [iii]  & $2.0\pm0.5$         & $1.6\pm0.3$         & $0.4\pm0.2$        & 36.9/50   \\    
3C~132             & 0.214   & 21.3 & [ii]   & $5_{-3}^{+4}$       & 1.7$^{\dag}$        & $22_{-8}^{+12}$  & 4.1/5 C   \\    
3C~153$^{\star}$   & 0.2769  & 16.2 & [i]    & -                   & 1.7$^{\dag}$        & $<1.5$           & -         \\    
3C~165             & 0.2957  & 19.4 & [ii]   & $3\pm 2$            & 1.7$^{\dag}$        & $23_{-8}^{+11}$  & 8.2/14 C  \\    
3C~166             & 0.2449  & 17.1 & [i]    & $<0.16$             & $1.6\pm 0.2$        & $80\pm 10$             & 20.1/19   \\    
3C~173.1           & 0.292   & 4.5  & [vi]   & $30_{-20}^{+200}$   & 1.7$^{\dag}$        & $27_{-16}^{+180}$ & 35.3/25 C \\    
3C~196.1           & 0.198   & 6.0  & [vii] & -                   & -                   & $<9.6$                                    & 20.5/13 C \\ 
3C~213.1           & 0.1939  & 2.4  & [i]    & $<0.43$             & $1.9_{-0.4}^{+0.5}$ & $4\pm 1$             & 15.4/9 C \\    
3C~236             & 0.1005  & 1.0  & [ii]   & $1.9_{-0.5}^{+0.6}$ & $1.4\pm 0.3$        & $12_{-4}^{+7}$  & 27.0/29   \\    
3C~288             & 0.246   & 0.8  & [vii] & -                   & -                   & $<4.7$                                    & 6.4/6     \\    
3C~310             & 0.0535  & 3.7  & [vii] & -                   & -                   & $<0.02$                                    & 10.5/17 C \\    
3C~326             & 0.0895  & 9.0  & [ii]   & $2.2_{-1.7}^{+2.8}$ & 1.7$^{\dag}$        & $0.2_{-0.1}^{+0.2}$        & 3.4/3 C   \\    
3C~349             & 0.205   & 1.9  & [ii]   & $0.9\pm 0.2$        & $1.4\pm 0.2$        & $60\pm10$        & 46.2/47   \\    
3C~353             & 0.0304  & 9.3  & [vi]   & $6.7_{-0.8}^{+0.9}$            & $1.7\pm 0.2$        & $3\pm1$              & 58.0/61   \\    
3C~357             & 0.1662  & 3.1  & [iii]  & $3\pm2$             & 2$\pm1$             & $22_{-12}^{+6}$ & 17/18 C   \\    
3C~388             & 0.0917  & 5.5  & [vii] &                   - & -                   & $<0.9$                                    & 62.5/59 C \\    
3C~401             & 0.2011  & 5.9  & [i]    & $<0.16$             & $1.7\pm 0.1$        & $5.0\pm 0.5$         & 15.9/18    \\    
3C~430$^{\star}$   & 0.0541  & 33.1 & [i]    & -                   & 1.7$^{\dag}$        & $<0.05$           & -         \\    
3C~460             & 0.268   & 4.72 & [iii]   & $25_{-11}^{+23}$    & 1.7$^{\dag}$        & $20\pm 10$             & 5.7/6     \\    
\hline

\multicolumn{8}{c}{\textbf{FRII-HERGs/BLRGs}}\\
3C~20              & 0.174   & 18.0 & [ii]   & $15_{-3}^{+4}$      & 1.7$^{\dag}$        & $110_{-20}^{+30}$  & 24.1/25 C \\    
3C~33 & 0.0596  & 3.4  & [v]    & $53_{-7}^{+8}$      & 1.7$^{\dag}$        & $100_{-20}^{+30}$    & 32.5/40   \\    
3C~61.1            & 0.184   & 7.9  & [iii]   & $29_{-12}^{+23}$    & 1.7$^{\dag}$        & $40_{-20}^{+40}$        & 30/18 C   \\    
3C~79              & 0.2559  & 8.7  & [iii]   & $33_{-10}^{+12}$     & 1.7$^{\dag}$        & $270_{-90}^{+130}$  & 14.4/16 C \\    
3C~98              & 0.0304  & 10.0 & [iv]   & $9.4_{-0.9}^{+1.0}$ & 1.7$^{\dag}$        & $5.3_{-0.3}^{+0.5}$  & 57.4/48   \\    
3C~105             & 0.089   & 12.0 & [iv]   & $43_{-6}^{+7}$      & 1.7$^{\dag}$        & $220_{-50}^{+90}$  & 13.6/12   \\ 
3C~133             & 0.2775  & 25.0 & [ii]   & $0.8_{-0.3}^{+0.4}$ & $2.0\pm 0.3$        & $190_{-50}^{+80}$  & 39.4/26   \\    
3C~135             & 0.1253  & 8.7  & [vi]   & $34_{-19}^{+32}$    & 1.7$^{\dag}$        & $14_{-8}^{+23}$  & 15.8/13 C \\    
3C~136.1$^{\star}$ & 0.064   & 32.0 & [i]    & -                   & 1.7$^{\dag}$        & $<0.06$            & -         \\    
3C~171        & 0.2384  & 5.7  & [ii]   & $7\pm1$      & $1.5\pm0.3$ & $130_{-50}^{+80}$  & 33.8/26   \\    
3C~180             & 0.22    & 14.0 & [iii]  & $70_{-50}^{+160}$   & 1.7$^{\dag}$        & $90_{-70}^{+1800}$      & 7.1/7 C   \\    
3C~184.1           & 0.1182  & 3.2  & [iii]   & $8\pm1$       & 1.7$^{\dag}$        & $110\pm 10$         & 19.8/24   \\    
3C~192             & 0.0598  & 3.9  & [iii]   & $34_{-7}^{+8}$     & $1.7\pm0.5$ & $2_{-1}^{+4}$      & 14.3/18  \\    
3C~223             & 0.1368  & 1.0  & [iii]   & $13_{-7}^{+13}$     & 1.7$^{\dag}$        & $20_{-8}^{+16}$  & 10/10 C    \\    
3C~223.1           & 0.107   & 1.3  & [ii]   & $28\pm6$      & 1.7$^{\dag}$        & $90_{-20}^{+30}$        & 9.3/12 C  \\   
3C~234             & 0.1848  & 1.8  & [vi]   & $17_{-6}^{+9}$      & 1.7$^{\dag}$        & $150_{-50}^{+70}$  & 6.6/8 \\ 
3C~277.3           & 0.0857  & 0.9  & [xiii]   & $27_{-5}^{+6}$      & 1.7$^{\dag}$        & $9_{-1}^{+2}$          & 23.1/21     \\    
3C~284             & 0.2394  & 0.9  & [i]    & $<0.91$             & $2.3\pm 1.0$        & $1.1_{-0.5}^{+0.4}$  & 1.4/5 C   \\    
3C~285             & 0.0794  & 1.3  & [vi]   & $38_{-6}^{+8}$      & 1.7$^{\dag}$        & $35_{-7}^{+10}$  & 7.7/11    \\    
3C~300             & 0.27    & 2.5  & [i]    & $<0.19$             & $1.4\pm 0.3$        & $13\pm 2$         & 12.7/10 C \\    
3C~303.1           & 0.267   & 3.0  & [ii]   & $18_{-16}^{+132}$   & 1.7$^{\dag}$        & $15_{-11}^{+400}$ & 0.1/2 C   \\    
3C~305             & 0.0416  & 1.3  & [viii]   & $<0.72$             & 1.7$^{\dag}$        & $0.04\pm0.01$             & 36.3/24 C   \\    
3C~321             & 0.096   & 3.8  & [ix]    & $26_{-13}^{+20}$    & 1.7$^{\dag}$        & $4_{-2}^{+4}$        & 61.2/40 C \\    
3C~327             & 0.1041  & 5.9  & [x]   & $30_{-18}^{+63}$    & 1.7$^{\dag}$        & $8_{-4}^{+31}$       & 46.6/25   \\    
3C~379.1           & 0.256   & 5.4  & [vi]   & $60_{-30}^{+70}$    & 1.7$^{\dag}$        & $110_{-70}^{+400}$  & 7.3/8 C   \\    
3C~381             & 0.1605  & 9.9  & [iii]   & $30_{-6}^{+7}$      & 1.7$^{\dag}$        & $240_{-50}^{+70}$  & 18.9/21 C \\    
3C~403             & 0.059   & 12.1 & [xi]  & $46\pm 3$           & 1.7$^{\dag}$        & $78_{-9}^{+10}$  & 51.5/57   \\    
3C~436             & 0.2145  & 6.7  & [iii]   & $48_{-15}^{+22}$    & 1.7$^{\dag}$        & $100_{-40}^{+80}$  & 14.4/15 C \\    
3C~452             & 0.0811  & 9.8  & [v]    &  $53_{-7}^{+8}$     & 1.7$^{\dag}$        & $100\pm20$  & 77.9/78   \\    
3C~456             & 0.233   & 3.7  & [ii]   & $7\pm1$             & 1.7$^{\dag}$        & $160\pm 20$         & 63.6/59 C \\    
3C~458             & 0.289   & 5.9  & [ii]   & $35_{-16}^{+20}$    & 1.7$^{\dag}$        & $150_{-70}^{+140}$  & 16.8/14 C \\    
3C~459             & 0.2199  & 5.2  & [xii] & $4_{-2}^{+3}$       & 1.7$^{\dag}$        & $12_{-2}^{+3}$         & 31.2/25 \\ \hline

\multicolumn{8}{l}{\scriptsize{$^{a}-$ All the adopted models are absorbed by the  Galactic column density:[i] po;[ii] zpha*po;[iii] zphabs*po+po; [iv] zphabs*(po+zgauss)}}\\
\multicolumn{8}{l}{\scriptsize{[v] zphabs*(po+zgauss)+po+pexrav); [vi] zphabs*(po+zgauss)+po;[vii] mekal;[viii] mekal+po;[ix] zphabs*(po+zgauss)+po+2zgauss;}}\\
\multicolumn{8}{l}{\scriptsize{[x] zphabs*(po+zgauss)+po+ mekal;[xi] zphabs*(po +zgauss)+po+2zgauss+mekal; [xii] zphabs*(po)+po+mekal; [xiii] zphabs*(po+zgauss)+zphabs*po}}\\
\multicolumn{8}{l}{\scriptsize{$^{b}-$ Statistics refers to the entire energy band assuming the model listed in column (4). ``C" indicates that the C-statistics was adopted.}} \\

\multicolumn{8}{l}{\scriptsize{$^{\dag}-$ fixed photon index}}\\
\multicolumn{8}{l}{\scriptsize{$^{\star}-$ luminosities estimated with PIMMS assuming a simple power law model with $\Gamma=1.7$}.}\\
\label{tabellone}
\end{tabular}
\end{center}
\end{table*}

\normalsize
\LTcapwidth=3.0\textwidth

\section{Discussion}
The aim of the present study is to explore the jet-accretion connection and the role of the environment in shaping the radio morphology in sources of different FR type.
Our X-ray results can be summarized as follows:
\begin{itemize}
        
    \item nearly 30\% of FRII-LERGs are in a  dense/extended gaseous environment, as attested by the \textit{Chandra} images. 
    Thermal gas is also detected in several images of the FRII-HERGs/BLRGs control sample. The extension of the emission seems to  
    suggest a galactic rather than an intergalactic origin;
    \item FRII-LERGs' spectra are generally well modeled by a power-law absorbed by a moderate intrinsic column density ($N_{\rm H} \sim$10$^{22}$~cm$^{-2}$). Conversely, FRII-HERGs/BLRGs have spectra rich in features and engines obscured by high column densities  ($N_{\rm H}\geq$10$^{23}$~cm$^{-2}$);
    \item FRII-LERGs are intrinsically less luminous than FRII-HERGs/BLRGs by a factor of ten in the 2$-$10~keV band.
    
\end{itemize}
\subsection{Are FRII-LERGs obscured FRII-HERGs/BLRGs?}
\label{sect_obscured}
The first scenario that we explore supposes that FRII-LERGs could be obscured FRII-HERGs/BLRGs.\\
Our analysis does not support this hyphotesis.
The gas column densities, estimated from the X-ray spectra, carry out information on the environment down to sub-parsec scales.
Looking at Table \ref{tabellone}, the difference between the FRII classes in terms of $N_{\rm H}$ is evident: FRII-HERGs/BLRGs are  more obscured than FRII-LERGs.
A two-sample test univariate program, TWOST \citep{Feigelson1985,Isobe1986}, that takes into account upper limits,  confirms that the  two samples are different, being  P$_{\rm TWOST}=7\times10^{-4}$.
We assume P$_{\rm TWOST}=0.05$ as the probability threshold to rule out the hypothesis that two samples are drawn from the same population. 

\begin{figure}
    \centering
        \includegraphics[width=8cm] {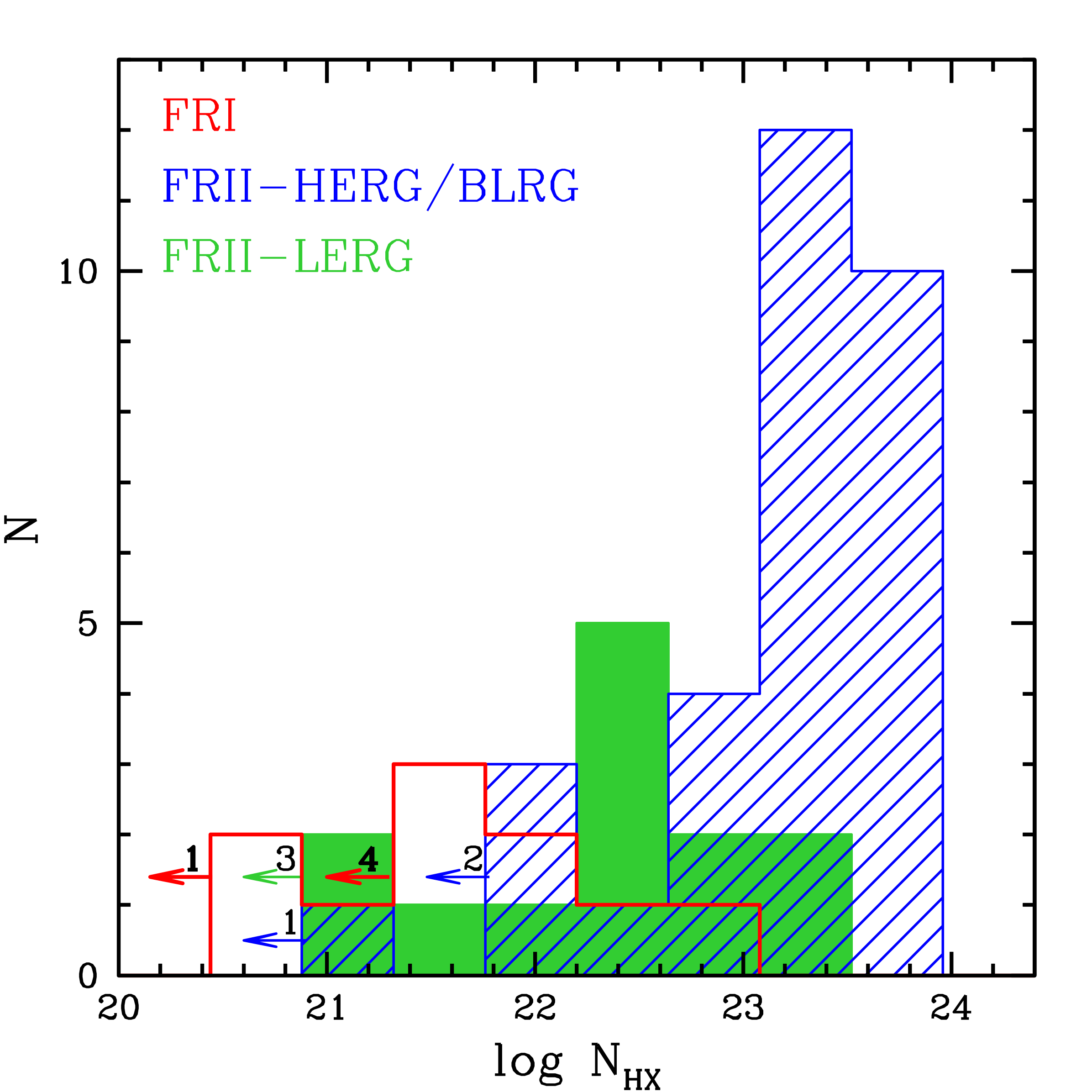}
    \caption{Distribution of the intrinsic gas column density $N_{\rm HX}$ in units of cm$^{-2}$, as measured in the X-ray band. The FRII-LERGs population (\textit{green}) is on average less obscured than  FRII-HERGs/BLRGs (\textit{blue}), but more than FRIs (\textit{red}). Arrows indicate upper limits with the number of sources per bin specified. FRIs data are taken from \protect\cite{Balmaverde2006}.}\label{img-nh} 
    \end{figure}

To take a step forward, 3CR/FRI radio galaxies with X-ray information available from literature \citep{Balmaverde2006} were also compared to the FRII samples. The peak of the $N_{\rm H}$ distribution in FRIs is clearly shifted to lower values, as shown in Figure \ref{img-nh}. There is a partial overlap with FRII-LERGs.  However, a TWOST test applied to FRII-LERGs and FRIs provides a probability of  P$_{\rm TWOST}=2\times10^{-2}$,  showing that these two samples are intrinsically different. Instead, the same test confirms that FRII-HERGs/BLRGs and FRIs are drawn from different populations (P$_{\rm TWOST}<10^{-4}$).
It is interesting to note that larger amounts of cold gas column densities are associated to radio galaxies with efficient accretion disks.

We conclude that there is an indication that the quantity of  obscuring matter (in the form of cold gas) is decreasing  from FRII-HERGs/BLRGs to  FRIs, with FRII-LERGs lying in between.\\

Another source of obscuration is the dust spread in the galaxy, that could affect our optical measurements. Note that the optical classification of radio galaxies provided by \cite{Buttiglione2009} is based on lines produced in the NLR.

The dust content can be estimated using the Balmer decrement. Adopting an extinction curve $\kappa(\lambda)$,  the intrinsic  color excess can be expressed as:

$$E(B-V)_i=\frac{2.5}{[\kappa(H_{\beta})-\kappa(H_{\alpha})]} \times \log\bigg[\frac{(H_{\alpha}/H_{\beta})_o}{3.1}\bigg]$$.

Details on the derivation of the above formula can be found in the Appendix of \cite{momcheva}.
The theoretical ($H_{\alpha}/H_{\beta}$) ratio is 2.86, as expected if the  temperature and the electron density of the NLR are  T=10$^4$~K and  $N_e$=10$^3$~cm$^{-3}$, respectively \citep{Osterbrock1989}. Actually, a value of 3.1 is considered the best prescription  for AGN \citep{Gaskell1982,Gaskell1984,Wysota1988,Tsvetanov1989,Heard2016}.
Several functional forms for the attenuation curve are present in literature.  The most used are: the Milky Way extinction curve  \citep{Cardelli1989}, the Large and Small Magellanic Cloud extinction curves  from \cite{Gordon2003} and a general extra-galactic extinction curve from \cite{Calzetti1997}. The reddening study was performed considering all the different  extinction curves. 
As the results are similar, hereafter the discussion is based on the  Milky Way  extinction curve.

\begin{figure}
    \centering
        \includegraphics[width=8cm] {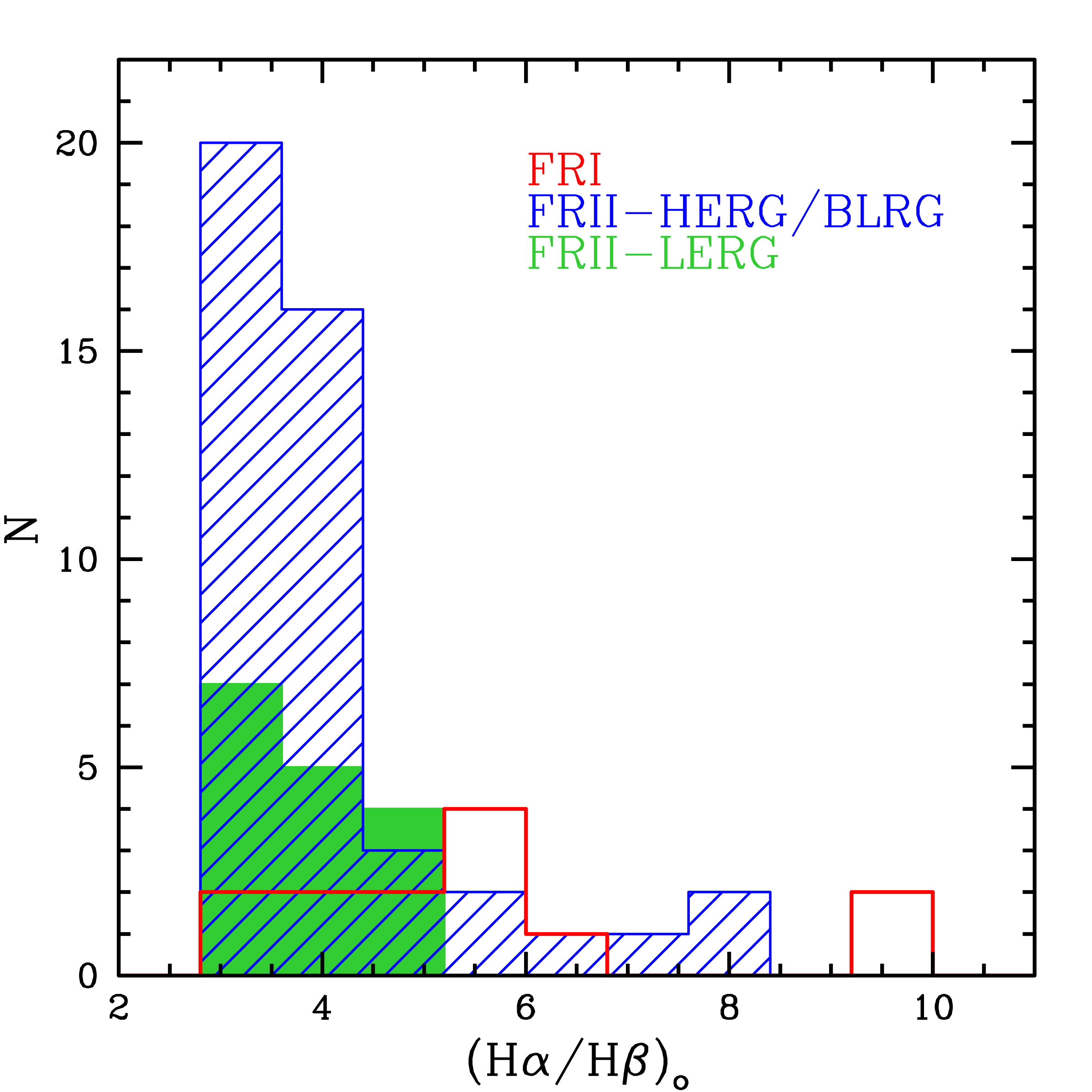}
    \caption{Distribution of the observed Balmer decrement (H$_{\alpha}$/H$_{\beta}$)$_{\rm o}$ used to estimate the dust extinction for the three classes of radio galaxies considered in this work, i.e.  FRII-LERGs, FRII-HERGs/BLRGs and FRIs. H$_{\alpha}$ and H$_{\beta}$ flux measurements are taken from \protect\cite{Buttiglione2009}. We assume the theoretical  (H$_{\alpha}$/H$_{\beta}$)$_{\rm o}$=3.1 as a reliable value for AGN (see Section 4.1). When (H$_{\alpha}$/H$_{\beta}$)$_{\rm o}<$3.1, the source is considered unabsorbed.}\label{img-ha_hb}
    \end{figure}

\cite{Buttiglione2010} provided the narrow H$_{\alpha}$ and $H_{\beta}$ fluxes  for the majority of 3C radio galaxies up to z=0.3. We could then investigate the  amount of dust in FRIs, FRII-LERGs and FRII-HERGs/BLRGs by simply comparing the $(H_{\alpha}/H_{\beta})_o$ ratio (assuming the same extinction curves for all the galaxies).
In Figure \ref{img-ha_hb}, a histogram of the  Balmer decrement for all the sources with detected lines (74 out of 79) is presented. When the flux ratio was less than the theoretical value, the source was considered unabsorbed (i.e. ($H_{\alpha}/H_{\beta})_o=3.1$). 
 It is immediate to note that the FRIs' distribution peaks to higher values of  $(H_{\alpha}/H_{\beta})_{o}$. Indeed, a  Kolmogorov-Smirnov (KS) test confirms that  FRIs are richer in dust than both FRII-LERGs (P$_{\rm KS}=0.01$) and FRII-HERGs/BLRGs (P$_{\rm KS}=0.008$). The current optical data do not allow us to exclude that the two FRII classes are drawn from the same population (P$_{\rm KS}=0.32$). 
 
 Therefore, the difference between FRII-LERGs and FRII-HERGs/BLRGs is intrinsic and not an artefact due to different absorbing screens.\\

\begin{figure}
    \centering
        \includegraphics[width=8.3cm] {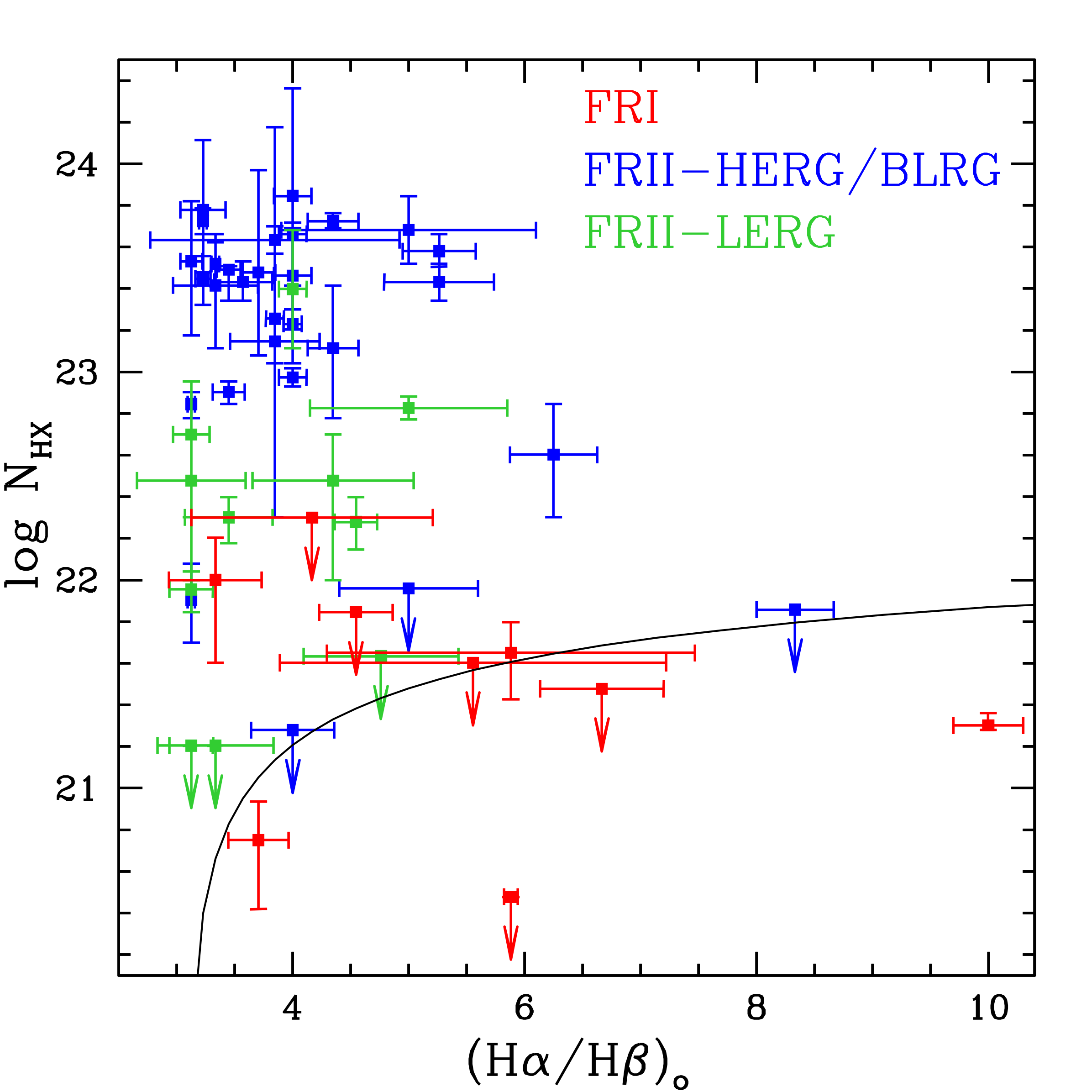}
    \caption{Column density ($N_{\rm HX}$), as obtained by the X-ray analysis, plotted versus the intrinsic reddening, as measured by the optical Balmer decrement $(H_{\alpha}$/$H_{\beta})_o$. $N_{\rm HX}$ decreases  from FRII-HERGs/BLRGs to FRI. FRII-LERGs occupy the middle region of the plot.  FRIs show the highest excess of colour and the lowest gas content. The black curve represents the expected $N_{\rm H}$ value assuming a dust-to-gas ratio $N_{\rm H}$/E(B-V)=$5.8\times 10^{21}$~atoms~cm$^{-2}$~mag$^{-1}$.}\label{img-ebmv-vs-nh}
    \end{figure}

In Figure \ref{img-ebmv-vs-nh} the column density ($N_{\rm HX}$)  measured in the X-ray band is plotted vs ($H_{\alpha}/H_{\beta})_o$. 
This plot traces the obscuring matter at different scales: $N_{\rm HX}$ maps the gas down to sub-pc scales, while the optical lines ($H_{\alpha}/H_{\beta})_o$ (i.e., E(B-V)), carries out information from the NLR.

Different classes appear to populate distinct regions of the plot: FRII-HERGs/BLRGs, having higher $N_{\rm HX}$ values,  mainly cluster in the upper part of the plot,  FRII-LERGs occupy a similar region but are shifted to lower $N_{\rm HX}$, FRIs lie at the bottom of $N_{\rm HX}$ but extend to ($H_{\alpha}/H_{\beta})_o$ up to 10.
Moreover, all FRIs are at the edge or below the N$_{\rm H}$ line that traces the expected amount of gas according to a standard Galactic gas to dust ratio  N$_{\rm H}$=$5.8\times 10^{21}$ E(B-V)~atoms~cm$^{-2}$~mag$^{-1}$ \citep{bohlin}.
Conversely, all the FRIIs are above the N$_{\rm H}$ line, suggesting a large amount of gas (although with different column densities) near the BH and paucity of dust in the NLR and/or along the galaxy.


The dichotomy between FRII-LERGs and FRII-HERGs/BLRGs is reinforced by the  X-ray analysis. The 
unabsorbed X-ray luminosity  divided by  the Eddington luminosity (L$_{\rm Edd}$=1.3$\times10^{38}$~M/M$_{\odot}$~erg~s$^{-1}$) is a direct proxy of the accretion rate (L$_{\rm 2-10~keV}$/L$_{\rm Edd}$; \citealt{Merloni2003}). The BH masses for the sources in our sample were calculated by exploiting the relation between the H-band host-galaxy magnitude (taken from \citealt{Buttiglione2009}) and M$_{\rm BH}$, provided by \cite{Marconi2003} (with a dispersion of $\sim$ 0.3~dex in the BH mass).
As expected, no significant difference in masses is observed among FRIs and FRIIs. The  M$_{\rm BH}$ range is narrow: $10^{8.5}-10^{9.5}$ M$_\odot$.

\begin{figure}
    \centering
       \includegraphics[width=6.9cm] {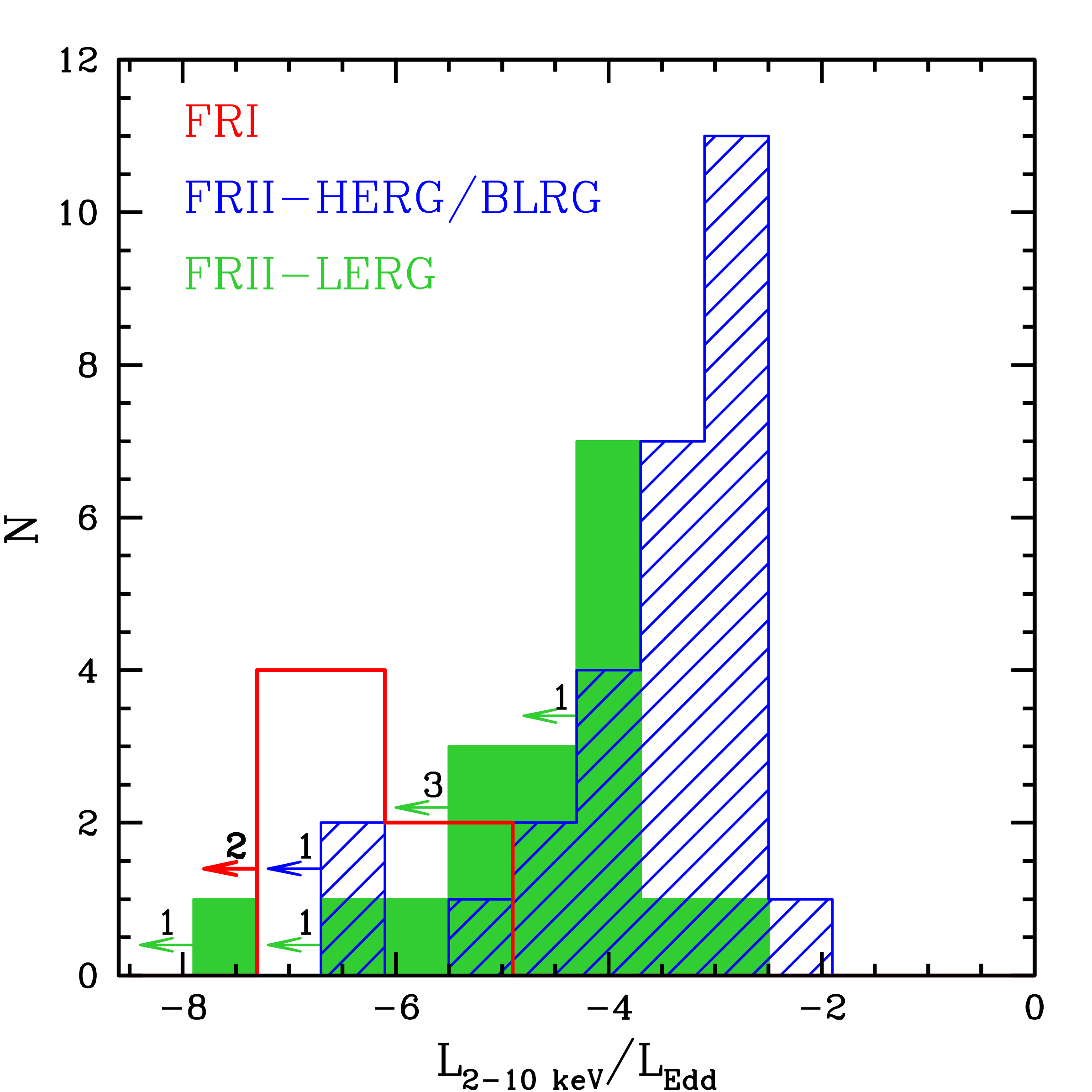}
        \includegraphics[width=6.9cm] {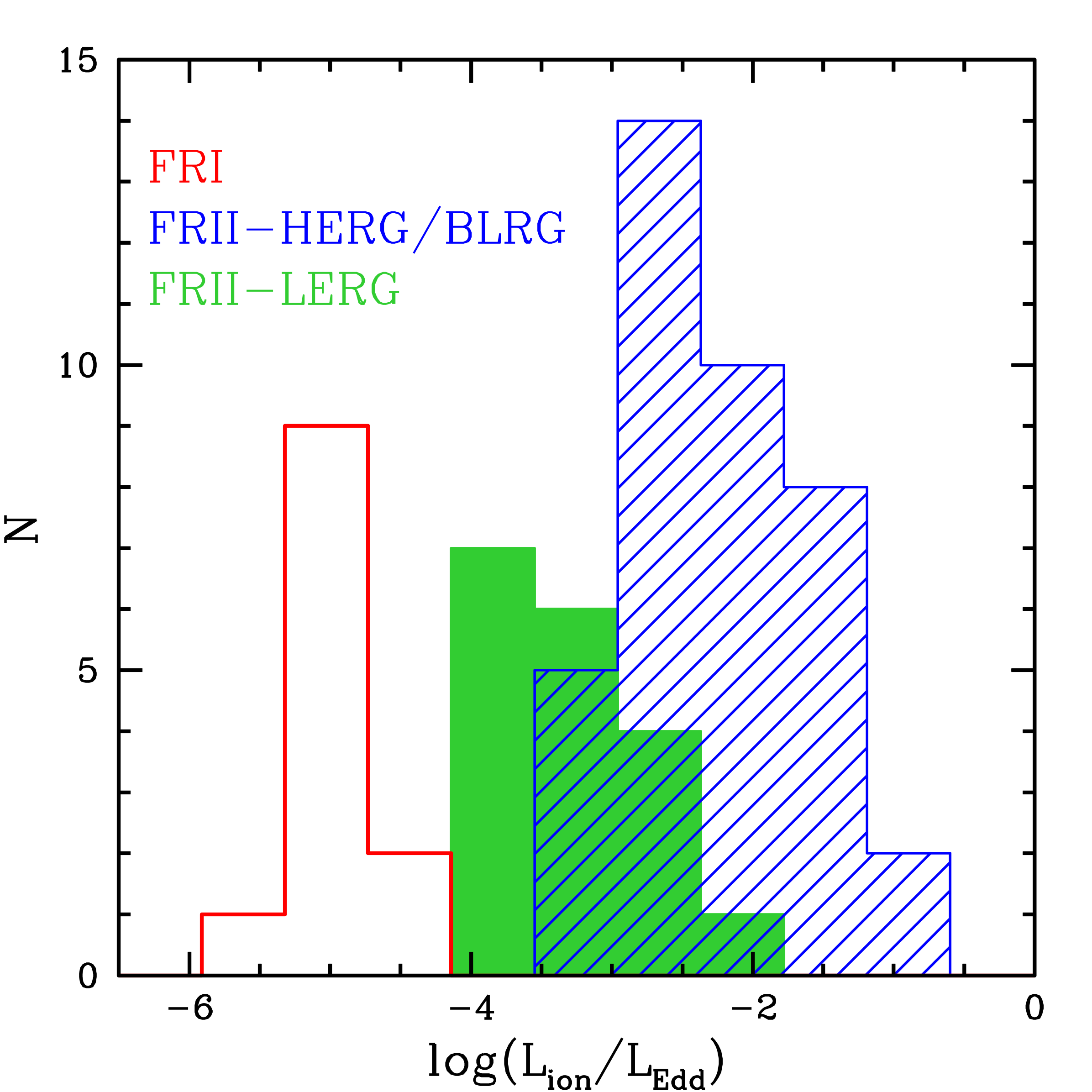}\qquad

    \caption{{\it Upper panel:} X-ray luminosity normalized for the Eddington luminosity. FRIs are in red, FRII-LERGs are in green and FRII-HERGs/BLRGs are in blue. Leftwards arrows indicate upper limits, with specified number per bin and population. {\it Lower panel:} estimated L$_{\rm ion}$/L$_{\rm Edd}$ ratio. Both accretion rate estimates show the same trend: FRIs at the lowest values, classical FRIIs at the highest ones, and FRII-LERGs in between.} \label{accretion}
    \end{figure}

The  upper panel of Figure \ref{accretion} shows the L$_{\rm 2-10~keV}$/L$_{\rm Edd}$ distribution for the three classes.
The distributions of FRIs and FRII-HERGs/BLRGs are clearly separated ($P_{TWOST}<10^{-4}$), while FRII-LERGs are in between. The displacement of FRII-LERGs' peak towards lower accretion rates is confirmed by a TWOST test, that associates a probability of $10^{-4}$ and 5.8$\times10^{-3}$ to the hypothesis that FRII-LERGs are drawn from the same 
parent population of FRII-HERGs/BLRGs and FRIs, respectively.
{\it Therefore, the nuclear activity is inherently different in FRIs, FRII-LERGs and FRII-HERGs/BLRGs.}

 Note that for RGs with ADAF-like engine the estimated X-ray luminosity could provide an upper limit of the accretion luminosity, as there could be a significant contribution from the jet emission. If this were the case, the separation between FRII-LERGS and FRII-HERGs/BLRGs would be even more pronounced.
Finally,  we note that the same trend is observed when 
the ionizing radiation L$_{\rm ion}$ in terms of the Eddington luminosity is considered (lower panel of Figure \ref{accretion}).
This quantity, defined as 
Log~L$_{\rm ion}\sim$ Log~$L_{[\rm OIII]}$ + 2.83 \citep{Buttiglione2009} is directly related to the accretion efficiency, being responsible for the excitation of the NLR gas.\\
The agreeement between upper and lower panels of Figure \ref{accretion} corroborates our previous conclusion based on the $(H_{\alpha}/H_{\beta})_o$ study.
As no dust correction was applayed to L$_{\rm ion}$, the different optical classification of FRIIs {\bf cannot} be ascribed to optical NLR obscuration.

\subsection{Were FRII-LERGs powerful FRII-HERGs/BLRGs in the past?}

The comparison of the X-ray and L$_{[OIII]}$ luminosities, accretion rates, and intrinsic nuclear absorption among the examined classes of sources have solidly established that FRII-LERGs have intermediate properties, lying between FRIs and FRII-HERGs/BLRGs. It is plausible that FRII-LERGs represent an evolutionary stage of FRII-HERGs/BLRGs.
A link between the accretion properties and the power of the produced jets is certainly expected, based on both theoretical arguments \citep[e.g.,][]{1977MNRAS.179..433B, 2007ApJ...658..815S} and observational works \citep[e.g.,][]{2006MNRAS.372...21A, Ghisellini2014}. 
Since the FRII-LERGs in this sample accrete at lower rates than classic FRII-HERGs/BLRGs, we would then expect their jets to be correspondingly less powerful. 
Contrarily to this expectations, their extended radio luminosities, generally assumed as  predictors of the total jet power \citep[e.g.,][]{2010ApJ...720.1066C}, are very similar (see Figure \ref{Pradio}).
This conflict can be bypassed considering the large-scale radio structures of FRII-LERGs as the heritage of a past AGN activity at higher efficiency. 
If the nuclear activity has recently decreased due, for instance, to the depletion of the cold gas reservoir, it is reasonable to think that this information may not have reached the large-scale radio structures yet, which are formed at kilo-parsec distances from the central engine.

The evolutionary scenario is also supported by a recent analysis of a large sample of low-redshift and low-luminosity FRIIs objects \citep{2017A&A...601A..81C}, which showed that the roughly one-to-one correspondence between FRII morphologies and powerful nuclei is not verified for this large population in the local universe (z<0.15). On the contrary, most of the FRIIs in the catalog compiled by \cite{2017A&A...601A..81C} are classified as LERGs. This could suggest that the local FRIIs are ``starved", i.e. they now miss the fueling cold material that made them shine in the past. 


\subsection{Does the environment play a role?}

Another  possibility is that the nuclei of FRII-LERGs, while not as powerful as in classic FRII-HERGs/BLRGs, can still form FRII morphologies due, for instance, to favorable environmental conditions. 
Several studies in literature identify the environment as the fundamental ingredient for the origin of the FRI/FRII dichotomy. 
\cite{Gendre2013},  studying the cluster richness for a large sample of radio galaxies, suggest that the relation between radio morphology and accretion mode is quite complex and attribute to the environment  an important role. They calculated the cluster richness (CR) following the method of \cite{Wing2011}, that is based on the count of galaxies  within a disk of 1~Mpc of radius around each analyzed target.
Then, they concluded  that FRII-HERGs/BLRGs and FRIs live in different ambients, being characterized by poor and rich environments, respectively. On the contrary, they found that the 29 FRII-LERGs of their sample can live both in clusters ($\approx 40\%$) and in scarcely populated regions (in terms of richness, $\approx 60\%$).

\begin{figure}
    \centering
        \includegraphics[width=7.5cm] {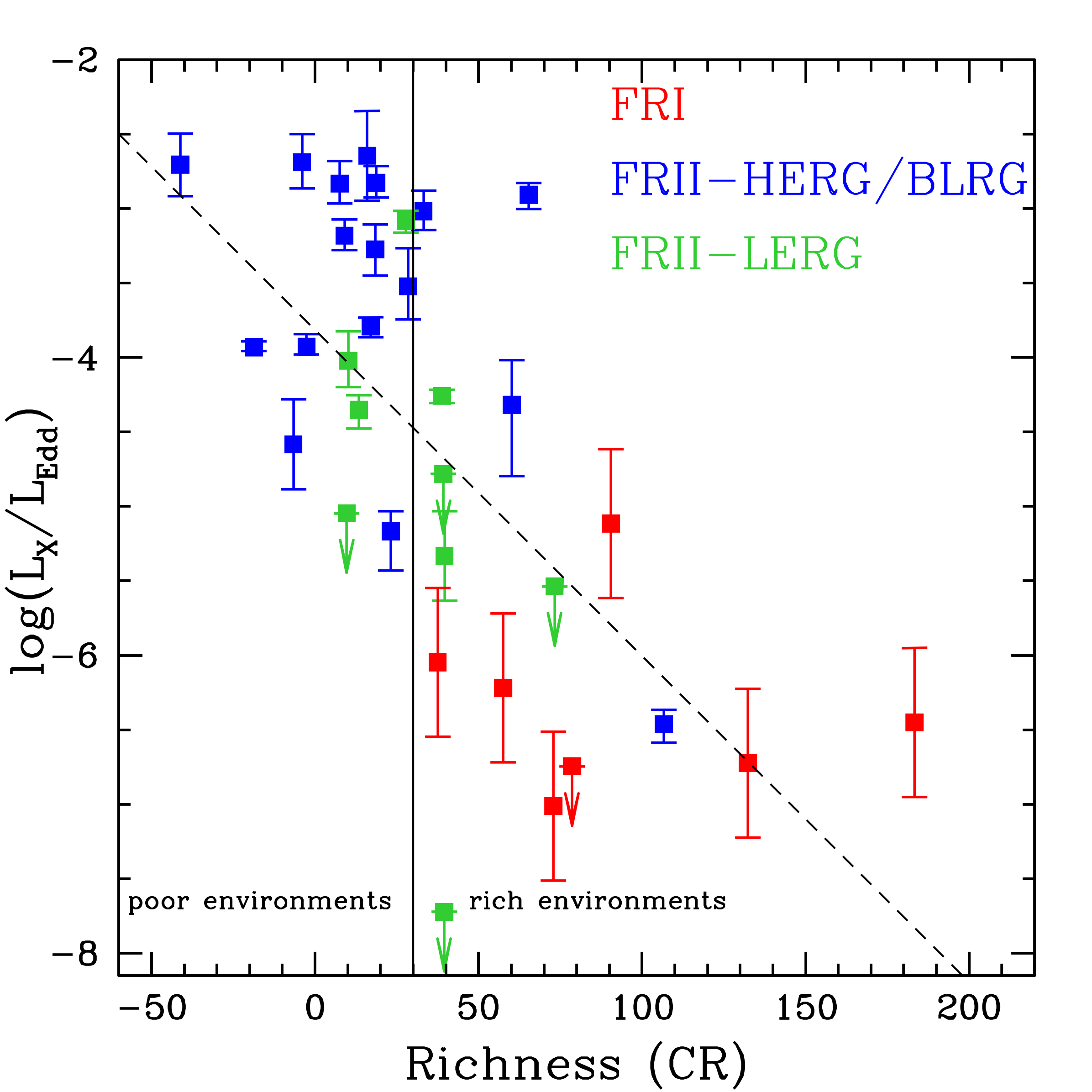}
    \caption{Environmental richness (CR) plotted versus the intrinsic Eddington-scaled X-ray luminosity. Color-coding is the same of  previous figures. Downwards arrows indicate upper limits. The solid black line is the threshold between poor and rich environments (CR=30), as indicated by \protect\cite{Gendre2013}. The dashed line represents the linear regression  of data computed with ASURV parametric EM algorithm, including upper limits. The equation is: Log(L$_X$/L$_{\rm Edd}$)=-$(0.022\pm 0.005)\times CR - (3.8\pm0.3)$.}\label{gendre}
    \end{figure}

Our X-ray data confirm that FRII-LERGs can actually be set in a dense and hot medium Indeed, 6 out of 19 are found in clusters  (see Figure \ref{fig-cluster}).. Unfortunately, our X-ray analysis, exploiting data from  public archives, suffers from an inhomogeneity of the exposure times that prevents a comparative study  on the environments among the different classes.  
Taking advantage of the richness study of \cite{Gendre2013}, we could  explore the relation between accretion, in terms of L$_{\rm 2-10~keV}$/L$_{\rm Edd}$, radio morphology and environment
for 18 FRII-HERGs/BLRGs, 8 FRIs and 9 FRII-LERGs.
In Figure \ref{gendre} the  X-ray luminosity scaled for the Eddington luminosity is plotted as a function of CR  for the three classes. The vertical line  is the limit between poor (CR$<$30) and rich (CR$>$30) environments proposed by \cite{Gendre2013}.
As expected, FRII-HERGs/BLRGs occupy the left upper corner (i.e. they are in less dense environments and have more efficient engines), while FRIs  are segregated in the right lower region (i.e. they have high CR and low L$_{\rm 2-10~keV}$/L$_{\rm Edd}$ values).
The intermediate accretion rates of FRII-LERGs put them in the middle part of the diagram, but, unlike the other classes, they fall into both sides of the threshold fixed at CR=30.
More impressive is the clear link in the whole sample between the richness of the environment and the accretion in terms of L$_{\rm 2-10~keV}$/L$_{\rm Edd}$. A Kendall$-\tau$ test in ASURV provides a very high probability that these quantities are correlated (P$_{{Kendall-\tau}}>99.9\%$).
With the appropriate caution  required  by  the  limited  number of sources,  this result suggests that the environment would have a strong impact on the accretion regime.  \\
The problem with this interpretation is that the radio luminosity of the extended components, related to the jet kinetic power, is similar in FRIIs, at odds with what is observed in  X-rays, where FRII-HERGs/BLRGs are brighter (more efficient accretors) than  FRII-LERGs. \\
However, the relation P$_{\rm jet}\propto L_{\rm 151~MHz}^{6/7}$, proposed by \cite{Willott1999} and widely used in literature, is calibrated on FRII radio galaxies and suffers from large uncertainties represented by a factor $f^{3/2}$, that mainly depends on the ratio between the energy in  protons and eletrons in the lobes. \cite{Willott1999} deduced that $f$ can span from 1 to 20,  implying a $P_{\rm jet}$  uncertainty of about two orders of magnitude. 
Other authors revisited this relation  measuring the jet power using the X-ray cavities produced by the interaction of FRI jets \citep{Birzan2008,Cavagnolo2010} or the hot-spot size and an equipartition magnetic field in FRIIs \citep{Godfrey2013}.
In particular, \cite{Cavagnolo2010} concluded that a k-value larger than 100 is necessary to match with the relation proposed by Willott.
Therefore, different classes of sources can require different values of f. We note that increasing f-values from FRIs to FRII-LERGs and FRII-HERGs/BLRGs could indeed mantain the proportionality between jet power and accretion luminosity. 

Indeed,  we note that the radio luminosity of a radio source can be amplified by radiative losses if the jet propagates through a dense environment. As claimed by \cite{Barthel1996} on the basis of a Cygnus A study, this effect would amplify the radio luminosity and, in turn, weakens its reliability as estimator of the AGN power.  


\begin{table}
\setlength{\arrayrulewidth}{0.3mm}
\caption{Spectral parameters of the soft X-ray component.}
\begin{center}

\begin{tabular}{p{0.08\textwidth}p{.1\textwidth}p{.08\textwidth}p{.1\textwidth}}
\hline
Name & $\Gamma_S$ & kT & $L_{0.5-2}$ \\
  && $(\rm keV)$ &(10$^{42}$~erg~s$^{-1}$)\\\hline
\multicolumn{4}{c}{\textbf{FRII-LERGs}}\\
 3C88              & 1.6=$\Gamma_{H}$ & -		      & $0.03_{0.005}^{0.007}$	      \\ 
3C173.1           & 1.7=$\Gamma_{H}$ & -		      & $0.3_{-0.2}^{+0.3}$	      \\    
3C196.1           & -		     & $3.2_{-0.8}^{+1.3}$    & $9\pm 1$	      \\    
3C288             & -		     & $3.7_{-1.2}^{+2.9}$    & $3.6_{-0.5}^{+0.6}$    \\    
3C310             & -		     & $1.0\pm0.2$	      & $0.04_{-0.009}^{+0.01}$   \\    
 3C353             & 1.7=$\Gamma_{H}$ & -		      & $0.04\pm0.01$   \\    
3C357             & 2=$\Gamma_{H}$   & -		      & $0.9_{-0.3}^{+0.7}$        \\    
3C388             & -		     & $2.1_{-0.2}^{+0.3}$    & $1.5_{-0.1}^{+0.2} $  \\    
3C460             & $1.7=\Gamma_{H}$ & -		      & $0.7\pm 0.3$	      \\    \hline
\multicolumn{4}{c}{\textbf{FRII-HERGs/BLRGs}}\\
3C33              & 1.7=$\Gamma_{H}$ & -		      & $1.0_{-0.2}^{+0.3}$   \\    
3C61.1            & 1.2$\pm$1.0      & -		      & $0.7\pm 0.4$	      \\    
3C79              & 1.7=$\Gamma_{H}$ & -		      & $1.8_{-0.7}^{+1.0}$   \\    
3C135             & $2.2\pm 1.0$     & -		      & $0.5\pm 0.2$	      \\    
3C180             & 1.7=$\Gamma_{H}$ & -		      & $0.9_{-0.4}^{+0.6}$   \\    
3C184.1           & 1.7=$\Gamma_{H}$ & -		      & $0.7\pm 0.3$	      \\    
 3C192             & 1.7=$\Gamma_{H}$ & -		      & $0.04\pm 0.01$	      \\    
3C223             & 1.7=$\Gamma_{H}$ & -		      & $1.4\pm 0.4$	      \\    
3C234             & $2.3\pm 0.5$     & -		      & $7\pm 1$	      \\    
 3C277.3           & 1.7=$\Gamma_{H}$ & -		      & $0.31_{-0.08}^{+0.1}$	      \\    
& \multicolumn{2}{l}{N$_{H,2}=1.1_{-0.4}^{+0.5~\ddag}$}&\\
3C285             & 1.7=$\Gamma_{H}$ & -		      & $0.14\pm 0.1$	      \\    
 3C305             & -		     & $0.8\pm0.2$	      & $0.013_{-0.005}^{+0.006}$	   \\    
3C321             & $2.8\pm 0.2$     & -		      & $0.8_{-0.06}^{+0.07}$   \\    
3C327             & 1.7=$\Gamma_{H}$ & $0.20_{-0.02}^{+0.04}$ & $1.9\pm 0.1$	      \\  
3C379.1           & 1.7=$\Gamma_{H}$ & -		      & $1.2_{-0.7}^{+0.9}$   \\    
3C381             & 1.7=$\Gamma_{H}$ & -		      & $2.\pm 0.5$	      \\    
3C403             & 1.7=$\Gamma_{H}$ & $0.20\pm 0.03$	      & $0.4\pm 0.04$	      \\ 
 3C436       & 1.7=$\Gamma_{H}$ & -	      & $0.4\pm 0.2$	      \\ 
3C452             & 1.7=$\Gamma_{H}$ & -		      & $0.1\pm0.04$   \\    
 3C459             & 1.7=$\Gamma_{H}$ & $0.7_{-0.1}^{+0.2}$    & $3.9_{-0.6}^{+0.7}$   \\\hline   
\multicolumn{4}{l}{\scriptsize{ $^{\ddag}-$ a secondary absorption component is also required for this }}\\
\multicolumn{4}{l}{\scriptsize{source (see \citealt{Worrall2016}).}}\\
\end{tabular}\label{soft_table}

\end{center}
\end{table}

\begin{table}
\setlength{\arrayrulewidth}{0.3mm}
\caption{Reprocessed Features.}
\begin{center}
    
\begin{tabular}{llc}
\hline
Name & Fe$K_{\alpha}$ line$^{\rm (a)}$ &Reflection (R)\\
  \hline
\multicolumn{3}{c}{\textbf{FRII-LERGs}}\\\\
 3C88               & 	unconstrained &-  	  \\
3C132              & EW$<949$				 &-  	  \\
3C165              & EW$<776$				 &-  	  \\
3C166              & EW$<388$				 &-  	  \\
3C173.1            & EW$\geq886$ 				 &-    \\
3C213.1            &  unconstrained			 &-    \\
3C236              & EW$<572$				 &-  	  \\
3C326              & unconstrained		 &-  	  \\
 3C349              & EW$<339$				 &-  	  \\
 3C353              & EW=$100\pm78$ 			 &- 	   \\ 
3C357              & EW$<610$				 &-  	  \\
 3C401              & unconstrained		 &-  	  \\
3C460              & EW$<359$				 &-  	  \\ \hline
\multicolumn{3}{c}{\textbf{FRII-HERGs/BLRGs}}\\\\			 
3C20              & EW$<281$				&-   	  \\ 
3C33              & EW=$139\pm89$           & R=$1.5_{-0.6}^{+0.4}$	      \\
3C61.1            & EW$<359$			&	-   	   \\ 
3C79              & EW$<157$			&	-   	   \\ 
3C98              & EW=$277\pm135$		&	-   	   \\ 
3C105             & EW=$178\pm132$		&	-   	   \\ 
3C133             & EW$<453$			&	-   	   \\
3C135             & EW$=916_{-719}^{+1474}$			&	-   	   \\ 
 3C171             & EW$<117$			&	-   	   \\ 
3C180             & EW$<744$			&	-   	   \\
3C184.1           & EW$<278$			&	-   	   \\ 
 3C192             & EW$<260$			&	-   	   \\ 
3C223             & EW$<836$			&	-   	   \\ 
3C223.1           & EW$<374$			&	-   	   \\ 
3C234             & EW$=900\pm400$			&	-   	   \\ 
 3C277.3           & EW$=200\pm100$			&	-   	   \\ 
3C284             & unconstrained	&	-   	   \\
3C285             & EW=$367_{-47}^{+144}$ &		-   	   \\ 
3C300             & unconstrained	&	-   	   \\
3C303.1             & unconstrained	&	-   	   \\
3C305             & unconstrained	&	-   	   \\
3C321             & EW=$988_{-474}^{+751}$ &	-	  	   \\ 
3C327             & EW=$2000_{-742}^{+3000}$&	-	   	   \\ 
3C379.1           & EW=$557_{-464}^{+1900}$	&	 -  	   \\ 
3C381             & EW$<2304$				 &  -	   \\ 
3C403             & EW=$153_{-15}^{+60}$ 	  & -	   \\ 
 3C436             & EW$<591$     &- 				\\
3C452             & EW=$172_{-65}^{+65}$  &R=2$_{-0.5}^{+0.4}$ \\
3C456             & EW$<156$			&	-   	   \\
3C458             & EW$<238$			&	-   	   \\
 3C459              & EW$<649$				 &-  	  \\
 \hline   
\multicolumn{3}{l}{\scriptsize{$^{\rm (a)}$ -- Observed iron line equivalent width in eV.}}\\
\label{reprocessed_table}
\end{tabular}
\end{center}
\end{table}

\normalsize

\section{Conclusions}

The comparison of the optical/X-ray luminosities, accretion rates, and intrinsic nuclear absorption between the three examined classes of sources (FRIs, FRII-LERGs, FRII-HERGs/BLRGs) have solidly established that FRII-LERGs have intermediate properties.
The measurement of moderate gas column densities in FRII-LERGs, combined with a modest dust reddening, enables us to directly reject the first discussed scenario (see Section \ref{sect_obscured}), in which FRII-LERGs are a highly obscured version of classical powerful FRIIs. 

Instead, the moderate N$_{\rm H}$ column densities and X-ray/[OIII] luminosities are indicative of a weak nuclear activity with respect to the more obscured FRII-HERGs/BLRGs.

This leads at least to two different interpretations: i) FRII-LERGs are ``starved"  classical FRII-HERGs/BLRGs, or ii) they are a separate class of radio galaxies with their own properties. 

In both cases, assuming a  P$_{\rm jet}-L_{\rm radio}$ relation,  FRII jets appear to carry out similar amount of energy independently on their optical classification (and different X-ray luminosities). This is difficult to explain within the  paradigm that assumes powerful jets produced by efficient accretors. 

In the former case, this apparent disagreement could be explained if  FRII-LERGs are switching from an efficient to an inefficient regime and this information may not have reached the lobe-scales yet. In the latter case, the different trend of radio (jet power) and X-ray luminosities (accretion power) can be reconciled if the usually adopted  P$_{\rm jet}-L_{\rm radio}$  relation does not properly take into account the jet interaction  with the surrounding medium. 

\section*{Acknowledgements}
The authors wish to thank the anonymous referee for constructive comments which helped to improve the paper.
DM and ET  acknowledge financial contribution from the agreement ASI-INAF n. 2017-14-H.O. This work is based on data from the \textit{Chandra} and XMM-\textit{Newton} Data Archive.
\label{lastpage}
\bibliography{paper}
\bibliographystyle{mnras}
\end{document}